\documentclass[nofootinbib,reprint,amssymb,superscriptaddress,twocolumn]{revtex4-2}
\usepackage[pdftex]{graphicx}
\usepackage{subfig}
\usepackage{hyperref}

\usepackage{caption}
\usepackage{ragged2e}
\DeclareCaptionFormat{justified}{\justifying#1#2#3}
\usepackage{tabularx}
\usepackage{tikz}
\usepackage{amsmath,amsfonts,amssymb,mathtools}
\usepackage{amsthm}

\newtheorem{corollary}{Corollary} 
\usepackage{tasks}
\usepackage{csquotes}
\usepackage{diagbox}
\usepackage{float}
\usepackage[utf8]{inputenc}
\usepackage{csquotes}
\usepackage{float}
\usepackage{textcomp}
\usepackage{bbm,bm}
\usepackage[normalem]{ulem}
\usepackage{comment}
\usepackage{physics}
\usepackage{xcolor}
\usepackage{CJKutf8}

\usepackage{pgfplots}
\definecolor{blueprl}{RGB}{46,48,146}
\usepgfplotslibrary{groupplots}

\usepackage{lipsum}

      \def\QUICS{Joint Center for Quantum Information and Computer Science,
NIST/University of Maryland, College Park, Maryland 20742, USA}
\def\JQI{Joint Quantum Institute, NIST/University of Maryland, College Park, Maryland 20742, USA}
\begin{document}
\title{Fault-Tolerant Heisenberg-Limited Quantum Sensing}
\author{Lorc{\'a}n O Conlon}
\email{lorcanconlon@gmail.com}
\affiliation{\JQI}
\affiliation{\QUICS}
\author{\begin{CJK}{UTF8}{gbsn}Yu-Xin Wang (王语馨)\end{CJK}}
\affiliation{\QUICS}
\author{Erfan~Abbasgholinejad}
\affiliation{\JQI}
\affiliation{\QUICS}
\author{Victor V. Albert}
\affiliation{\QUICS}
\author{Michael J. Gullans}
\affiliation{\QUICS}
\author{Alexey~V.~Gorshkov}
\affiliation{\JQI}
\affiliation{\QUICS}

\begin{abstract}
Quantum sensors hold great promise for achieving better sensitivity in the measurement of physical quantities compared to their classical counterparts. However, the conditions under which quantum advantage in sensing can be achieved are rather restrictive, and most quantum enhancements in sensing are lost in the presence of noise, errors, or a poorly calibrated system. To overcome these limitations, we are motivated to import ideas from fault-tolerant quantum computing to quantum sensing. Specifically, we consider a qubit noise model where the probability of phase-flip errors is exponentially smaller (in qubit number) compared to the probability of bit-flip errors that occur with probability $p$. For this noise structure, we demonstrate that, given a total sensing time $T$, Heisenberg scaling can be attained for times up to $T\propto 1/p^{(N+1)/2}$ for a $N$-qubit repetition code, in contrast with $T\propto 1/p$ without using a fault-tolerant sensing protocol. 
\end{abstract}
\maketitle

\section{Introduction}
Quantum sensing aims to use quantum mechanical resources such as quantum superposition, entanglement and squeezing to estimate properties of a system better than is possible using only classical resources~\cite{giovannetti2011advances}. Quantum sensing has already enabled advantages in gravitational wave detection~\cite{aasi2013enhanced,tse2019quantum} and biosensing~\cite{casacio2021quantum}, with several other promising avenues. One of the most commonly explored scenarios for entanglement advantage in sensing is estimating a rotation about the $Z$-axis of a $N$-qubit system~\cite{bollinger1996optimal}, i.e.~predicting $\phi$ given access to the Hamiltonian $H=\frac{1}{2}\phi\sum_iZ_i$ for unit time, where $Z_i$ is the Pauli-$Z$ operator acting on the $i$th qubit. In this setting, the standard quantum limit (SQL), corresponding to using unentangled probe states $\ket{\psi}=\ket{+}^{\otimes N}$, is given by $\sigma_\phi\geq1/\sqrt{N\nu}$, where $\nu$ is the number of experimental repetitions, and $\sigma_\phi$ is the standard deviation in our estimate of $\phi$. In contrast, when using a $N$-qubit Greenberger–Horne–Zeilinger (GHZ) state~\cite{greenberger1989going}, $\ket{\psi}=(\ket{0}^{\otimes N}+\ket{1}^{\otimes N})/\sqrt{2}$, the Heisenberg limit can be achieved, $\sigma_\phi\geq1/(N\sqrt{\nu})$, a factor $\sqrt{N}$ improvement~\cite{giovannetti2011advances}. If the $N$-qubit rotation channel can be accessed sequentially rather than in parallel, then the Heisenberg limit can be achieved without entanglement. That is, given $N$ sequential uses of $H=\phi Z_1/2$ for unit time (and then repeating this whole process $\nu$ times) with the state $\ket{\psi}=\ket{+}$, this single-qubit sensor can also achieve the Heisenberg limit in $N$, $\sigma_\phi=1/(N\sqrt{\nu})$.

Given this potential advantage, there is great interest in performing quantum sensing with large numbers of entangled qubits and / or while maintaining coherence over large timescales. However, most demonstrations of entanglement-enhanced sensing are limited in scale, employing only a small number of photons~\cite{daryanoosh2018experimental,nielsen2023deterministic} or qubits~\cite{marciniak2022optimal,conlon2023approaching,conlon2023discriminating,yung2025saturating,yung2026beating}. The reason for this is well established---noise and system imperfections rapidly degrade quantum sensing performance~\cite{Huelga1997,dorner2009optimal,Escher2011,demkowicz2012elusive}. Crucially, for larger $N$, this degradation happens faster. For example, consider the setting whereby, in addition to the rotation above, the $N$-qubit GHZ state is subject to $Z_i$ errors that occur on each qubit independently with probability $p$. In this setting the GHZ coherence will decay as $(1-2p)^N\approx e^{-2pN}$. Thus the visibility of the interference fringe (and hence the information about $\phi$) is suppressed exponentially in $Np$. From this, we can determine that to preserve Heisenberg scaling in $N$ we require $p$ to scale approximately as $1/N$---entirely unfeasible in practice.

This extreme fragility to noise at first glance appears to significantly limit the regime in which entanglement-enhanced quantum sensing can be useful. However, this pessimism was displaced by several proposals that applied quantum error correction to restore Heisenberg scaling in the presence of certain noise models~\cite{ozeri2013heisenberg,kessler2014quantum,Arrad2014,dur2014improved}. These proposals were later unified into a framework for determining when quantum error correction can restore Heisenberg scaling for a unitary evolution embedded in a noisy channel~\cite{sekatski2017quantum,demkowicz2017adaptive,zhou2018achieving,zhou2021asymptotic}. These conditions are related to the geometry of the noisy channel: for channels where the dynamics are described by a Lindblad master equation with Hamiltonian $H$ and jump operators $\{L_i\}$, the condition is known as Hamiltonian not in Lindblad span (HNLS)~\cite{zhou2018achieving}. Similarly, for a channel modeled using Kraus operators, the corresponding condition is known~\cite{zhou2021asymptotic}. 
While these conditions represent a significant advance in quantum sensing and have even led to some initial experimental demonstrations~\cite{unden2016quantum,niroula2024quantum}, there remain many practical issues impeding the widespread application of error-corrected quantum sensing~\cite{rojkov2022bias}. A significant practical issue is that existing error-corrected sensing works focus on noise in the quantum channel exclusively. All other operations including state preparation, syndrome extraction, and measurements are assumed to be perfect. 




For quantum sensors to be truly scalable, they need to overcome these limitations. A natural approach is to apply ideas from fault-tolerant quantum computing to quantum sensing~\cite{shor1996fault,preskill1998fault}. This motivation comes from the fact that there exist error correcting codes that correct arbitrary single-qubit errors~\cite{laflamme1996perfect} (potentially removing the difficulty of small perturbations in the noise model) and that by definition errors in one part of the circuit do not propagate to cause logical errors (potentially removing the requirement of perfect state preparation, syndrome extraction, and measurements). Recent experimental progress towards fault-tolerant quantum computation provides further hope that these advantages may be realized in the near-term~\cite{postler2022demonstration,bluvstein2024logical,google2025quantum}. However, as we shall discuss, known no-go theorems prevent a naive application of fault tolerance to quantum sensing. In this work, we examine how the ideas of fault tolerance can be applied to quantum sensing in spite of these no-go theorems. We demonstrate that, under a restricted noise model, Heisenberg scaling in total time can be achieved through the use of a simple repetition code, even with errors at every point in the circuit. These results contribute to ongoing efforts to find scalable methods of achieving quantum enhanced sensing, overcoming some of the many limitations outlined above.


This work proceeds as follows. In Sec.~\ref{sec:intro}, we present the preliminary material introducing the basics of quantum sensing, error correction, and fault tolerance. In Sec.~\ref{sec:basicFT}, we present basic results regarding fault-tolerant quantum sensing. Then, in Sec.~\ref{sec:FTlimitednoise}, we present our main result---under a specific noise model, fault-tolerant sensing extends the time-scale over which Heisenberg scaling can be achieved. Finally, in Sec.~\ref{sec:conclusion}, we conclude and present ideas for future work.


Before commencing, we note that during preparation of this manuscript, we became aware of a recent result on fault-tolerant quantum sensing~\cite{sahu2026achieving}. We compare our work to this study in Sec.~\ref{sec:comparison} below.

\begin{figure}[hbtp]
    \centering
\includegraphics[width=0.99\columnwidth]{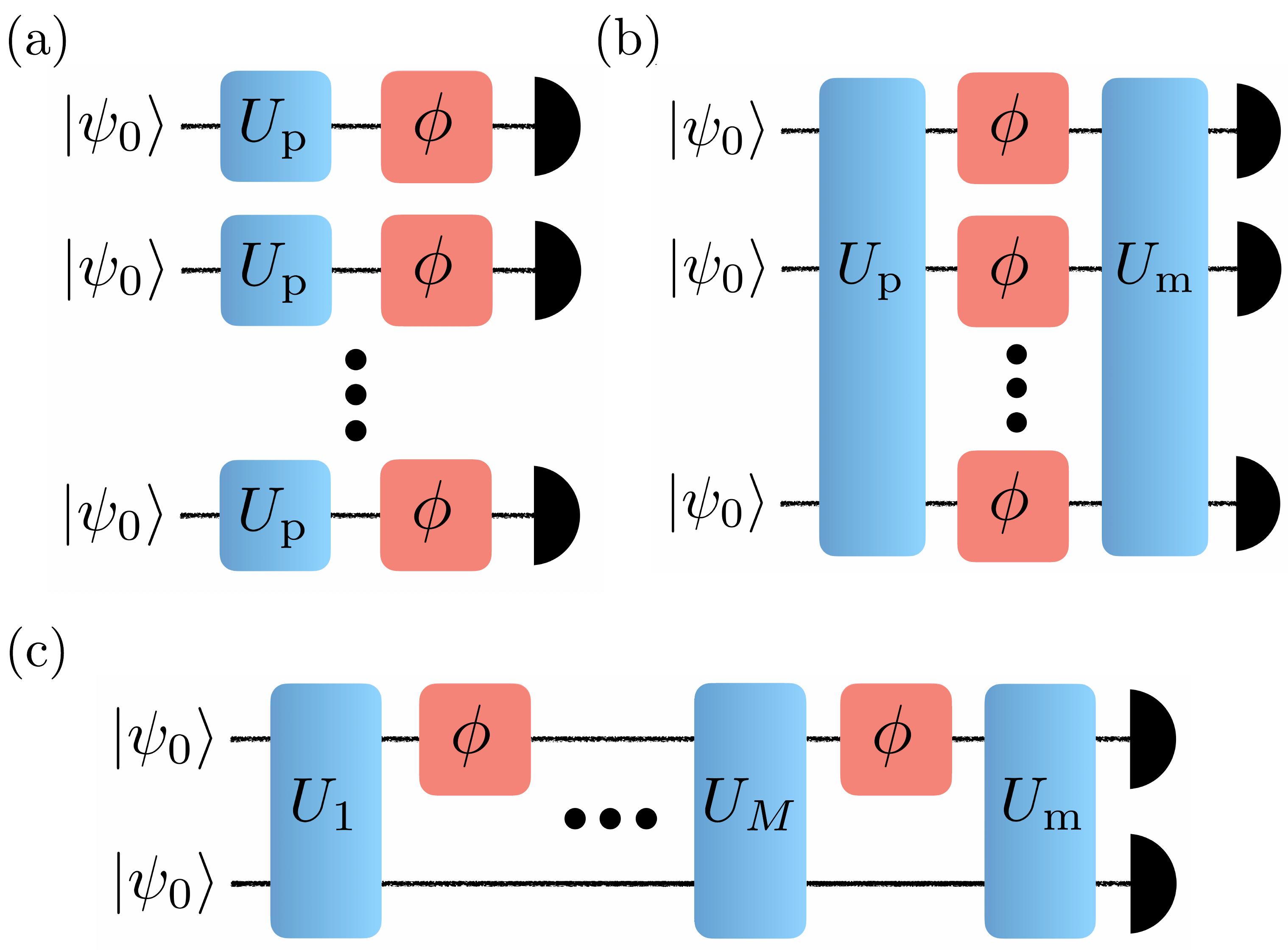}
    \caption{\textbf{Hierarchy of sensing protocols.} (a) Separable sensing, where each quantum channel encoding the parameter of interest $\phi$, represented by the red box, is probed with a separable state. $U_p$ represents a unitary preparing some quantum state and $\ket{\psi_0}$ may include ancilla qubits. Given $N$ uses of the same channel, the variance in estimating $\phi$ will scale as $\sigma_\phi^2\geq1/N$.   
    (b) Entangled sensing where $U_p$ acts across all $N$ qubits (and possibly ancillas) to prepare some optimal (potentially entangled) state for sensing $\phi$. An optimal measurement across all $N$ qubits (and possibly ancillas) is performed, represented by $U_m$, and Heisenberg scaling is achieved $\sigma_\phi^2\geq1/N^2$. (c) $M$ sequential uses of the rotation channel are allowed. Before each channel, a control operation $U_i$ is applied, and at the end an optimal measurement is performed across both the probe and ancilla, represented by $U_m$. This is termed full and fast control in Ref.~\cite{sekatski2017quantum}, and Heisenberg scaling is achieved $\sigma_\phi^2\geq1/M^2$. When $M=N$, all of the protocols access the rotation channel the same number of times. Note that in all of the above settings we could consider that each channel is accessed for unit time. In this way, Heisenberg or SQL scaling in total time $T=M$ can be obtained.}
    \label{differentQuantumSensing}
\end{figure}

\section{Preliminary material}
\label{sec:intro}
Here we set out all the pre-requisite information for understanding our results. We begin in Sec.~\ref{subsecQPE} by introducing the main theoretical tools of quantum sensing. In Sec.~\ref{subsecEC}, we then introduce the basics of error correction, before describing fault tolerance in Sec.~\ref{subsecFT}

\subsection{Quantum parameter estimation}
\label{subsecQPE}
In this work, we focus on the estimation of a unitary parameter. There are several commonly considered settings in quantum metrology. The first setting is that shown in Fig.~\ref{differentQuantumSensing} (a) whereby $N$ separate states are prepared and allowed to interact with a Hamiltonian of the form \begin{equation}
\label{eq:Hamiltonian}
    H=\phi h_0=\phi\sum_{i=1}^{N}\sigma_{\bar{n}}^{(i)}/2\;,
\end{equation}
for unit time, where $\sigma_{\bar{n}}^{(i)}$ denotes some local operator acting on the $i$th qubit, and we use subscript $\bar{n}$ to denote the direction of the encoded signal. Although $\bar{n}$ can in principle point along any direction in the Bloch sphere, for simplicity, and without loss of generality, here we focus on the case where $\sigma_{\bar{n}}^{(i)}=Z_i$ for all $i$. In the second setting, shown in Fig.~\ref{differentQuantumSensing} (b), a---possibly entangled---$N$ qubit state is prepared and allowed to interact with the same Hamiltonian $H$, also for unit time. A third setting, shown in Fig.~\ref{differentQuantumSensing} (c), is to allow a state $\ket{\psi}$ to evolve under the Hamiltonian $H=\phi Z/2$ for unit time, repeated $M$ times sequentially with control operations in between each channel use, such that the total evolution time is $T=M$. In this way, when $M=N$ the total number of channel uses (resources) is the same in Fig.~\ref{differentQuantumSensing} (a), (b) and (c). Here, and throughout, we use $M$ for the number of channel uses to avoid confusion with qubit number $N$. 

The central aim in quantum metrology is to minimize the mean squared error (MSE) in our estimate of $\phi$ for a given resource (number of channel uses), defined as
\begin{equation}
    \text{MSE}=\mathbb{E}[(\phi-\hat{\phi})^2]\;,
\end{equation}
where $\hat{\phi}$ is the predicted value of $\phi$. In this setting, it is known that a fundamental limit on the minimum error for estimating $\phi$ is~\cite{boixo2007generalized}
\begin{equation}
    \sigma_\phi\geq\frac{1}{\sqrt{\nu}||h_0||T}\;,
\end{equation}
where $\sigma_\phi^2=$MSE (assuming an unbiased estimator, see below), 
the factor of $\sqrt{\nu}$ accounts for $\nu$ repetitions of the experiment, and  $||h_0||$ denotes the semi-norm of $h_0$. Therefore, any protocol that saturates this bound is provably optimal.

For the Hamiltonian in Eq.~\eqref{eq:Hamiltonian}, we have $||h_0||=N$, recovering the Heisenberg limit discussed in the introduction
\begin{equation}
    \sigma_\phi\geq\frac{1}{TN\sqrt{\nu}}\;.
\end{equation}
The total evolution time in Fig.~\ref{differentQuantumSensing} (b) is $T=1$. Hence, in this setting, the above bound can be achieved by a GHZ state. If instead the experimenter used the state $\ket{+}$ and interacted with the Hamiltonian $M$ times sequentially, each for unit time, [as in Fig.~\ref{differentQuantumSensing} (c)], the state would evolve to $(\mathrm{e}^{-iM\phi/2}\ket{0}+\mathrm{e}^{iM\phi/2}\ket{1})/\sqrt{2}$. Measuring in the $X$-basis also gives an estimator that achieves the Heisenberg limit of $\sigma_\phi=1/(M\sqrt{\nu})$. In contrast, when restricted to not using entanglement or sequential interactions, i.e. probing the Hamiltonian $\phi h_0$ for unit time using the state $\ket{+}^{\otimes N}$, as in Fig.~\ref{differentQuantumSensing} (a), one recovers the SQL
\begin{equation}
        \sigma_\phi\geq\frac{1}{\sqrt{N\nu}}\;.
\end{equation}
We note that considering evolution for unit time does not affect the problem, as this assumption can be removed simply by rescaling $\phi$. This allows a simple comparison of the total resources used in parallel and sequential sensing. Going forward, we shall drop the explicit dependence on $\nu$, unless it is required for a fair accounting of resources used.

\subsubsection{Classical Fisher information}
A useful quantity for bounding the precision achieved in parameter estimation is the classical Fisher information (CFI) $\mathcal{J}$. For any complete description of an experiment, the CFI provides a bound on the MSE as
\begin{equation}
\label{eq:CFI}
    \sigma_\phi\geq\frac{1}{\sqrt{\mathcal{J}}}\;.
\end{equation}
 Given the quantum state after passing through the channel $\rho(\phi)$, the experimenter will perform a measurement described by a POVM which is a set of positive operators $\{\tilde{\Pi}_k\}$ that sum to the $D$-dimensional identity operator, $\sum_{k}\tilde{\Pi}_{k}=I_D$. The $k$-th measurement outcome occurs with probability
\mbox{$p_k=\text{Tr}[\rho\tilde{\Pi}_{k}]$}. Based on the observed measurement outcomes, the experimenter arrives at an estimated value of $\phi$, $\hat{\phi}$. For locally unbiased estimation around a known point, $\phi_0$, we require~\cite{fujiwara2006strong,holevo2011probabilistic}
\begin{equation}
\begin{split}
 &    \langle\hat{\phi}\rangle=\phi_0\\
&      \left.\frac{\partial\langle\hat{\phi}\rangle}{\partial\phi}\right|_{\phi=\phi_0}=1\;.
\end{split}
\end{equation}
In this setting, the CFI can be computed as
\begin{equation}
\label{eq:CFIdefin}
\mathcal{J}=\mathbb{E}\left[\left(\frac{\partial \log (P(\phi))}{\partial \phi}\right)^2\right],
\end{equation}
where $P(\phi)$ is the probability distribution of potential outcomes $x_k$ occurring with probability $p_k$. For locally unbiased estimators, Eq.~\eqref{eq:CFI} holds. 

\subsubsection{Quantum Fisher information}
The CFI, although useful, depends on the measurement chosen by the experimenter. The quantum Fisher information (QFI) $\mathcal{J}_\text{S}$ is a generalization of the CFI that does not require a specified measurement to be computed. Given a quantum state $\rho(\phi)$, one can compute the symmetric logarithmic derivative (SLD) operator $\mathcal{L}_\phi$ as
\begin{equation}
    \frac{\partial\rho}{\partial\phi}=\frac{1}{2}(\rho \mathcal{L}_\phi+\mathcal{L}_\phi\rho)\;.
\end{equation}
From this the QFI can be computed as
\begin{equation}
    \mathcal{J}_\text{S}=\text{Tr}[\rho \mathcal{L}_\phi \mathcal{L}_\phi]\;.
\end{equation}
The QFI then bounds the MSE as
\begin{equation}
    \sigma_\phi\geq\frac{1}{\sqrt{\mathcal{J}_\text{S}}}\;.
\end{equation}
 This quantity can be very useful, as it allows computation of the fundamental precision limit without having to optimize over all possible measurements. For single parameter estimation as considered here, the bound is tight, i.e. $\mathcal{J}_\text{S}=\max_{\tilde{\Pi}_i}\mathcal{J}$.

\subsection{Error Correction}
\label{subsecEC}

QEC provides a systematic framework for protecting fragile quantum information by encoding it redundantly into a larger Hilbert space. In its most familiar incarnation, QEC was developed to enable reliable quantum computation in the presence of noise: rather than attempting to engineer an effectively noiseless device at the physical layer, one instead groups the dominant error processes into a correctable set and repeatedly removes them using carefully designed measurements and feedback \cite{Shor1995,Steane1996,Gottesman1997,KnillLaflamme1997}. In the context of quantum sensing, the same philosophy is appealing for a complementary reason: as in Fig.~\ref{differentQuantumSensing} (b) and (c), the quantum enhancement in sensing is ultimately tied to how long coherent phase information can be maintained while the probe interacts with the signal or how large the probe state can be while maintaining coherence. Any method that extends the usable coherence time or number of coherent qubits---without erasing the signal itself---can potentially translate into improved sensitivity.

We model noise on a physical probe system by a completely positive trace-preserving map $\mathcal{E}$ acting on states $\rho$,
\begin{equation}
\label{eq:Krausevol}
  \mathcal{E}(\rho)=\sum_{a} K_a \rho K_a^\dagger,
\end{equation}
where $\{K_a\}$ are Kraus operators representing the error processes that satisfy $\sum_a K_a^\dagger K_a = I_D$. A code is specified by an isometric encoding $V:\mathcal{H}_L\to \mathcal{H}_P$ from a logical space $\mathcal{H}_L$ into a physical space $\mathcal{H}_P$, with code projector $\Pi_C = VV^\dagger$. We seek a recovery channel $\mathcal{R}$ such that $\mathcal{R}\circ \mathcal{E}$ acts as identity on all encoded states
\begin{equation}
  (\mathcal{R}\circ \mathcal{E})(V\rho_L V^\dagger)= V\rho_L V^\dagger \;,
\end{equation}
for all $\rho_L$ that are density matrices on $\mathcal{H}_L$. The Knill--Laflamme conditions characterize when a set of errors $\{K_a\}$ is perfectly correctable on the code space:
\begin{equation}\label{eq:KL}
  \Pi_C K_a^\dagger K_b \Pi_C = \alpha_{ab}\,\Pi_C \qquad \text{for all } a,b,
\end{equation}
for some Hermitian matrix $\alpha$ \cite{KnillLaflamme1997}. Intuitively, Eq.~\eqref{eq:KL} says that distinct correctable errors map the code space into mutually distinguishable subspaces without leaking logical information into the error syndrome. This is the mathematical origin of a typical QEC cycle: (i) encode logical information, (ii) extract an error syndrome via measurements that preserve $\Pi_C$, and (iii) apply a correction conditioned on the syndrome.


\subsubsection{Error-corrected quantum metrology}
As discussed in the introduction, although entanglement or long coherence times can achieve the Heisenberg limit in the ideal setting, for many realistic noise models this scaling is lost. In these settings, even with optimized measurements and perfect adaptive control,  the MSE in estimating $\phi$ will scale as the SQL \cite{Huelga1997,Escher2011,demkowicz2012elusive}. This motivates incorporating active protection into sensing protocols: rather than treating noise as an unavoidable limitation, one attempts to continuously suppress it while preserving the information of interest. In the context of error-corrected quantum metrology, it is useful to describe sensing problems as a continuous time evolution. For this setting, we consider a parameter $\phi$ encoded through a Hamiltonian $H(\phi)=\phi h_0$ acting on a probe system, while the probe is simultaneously subject to Markovian noise generated by Lindblad operators $\{L_k\}$,
\begin{equation}\label{eq:lindblad_metrology}
  \frac{d\rho}{dt} = -i[ \phi h_0,\rho] + \sum_k\Big(L_k\rho L_k^\dagger - \tfrac12\{L_k^\dagger L_k,\rho\}\Big).
\end{equation}
Note that, as will be discussed below, for sufficiently small $\delta t$, the jump operators can form an approximate Kraus operator representation [as in Eq.~\eqref{eq:Krausevol}] for this type of evolution accurate to $O(\delta t^2)$. This framework introduces a central tension that is specific to error-corrected quantum metrology: one must remove the effect of the $L_k$ without also removing the Hamiltonian imprint of $\phi$. Put differently, the goal is not necessarily to preserve an arbitrary logical state, but rather to preserve the distinguishability between nearby parameter values. 


A standard error-corrected sensing procedure discretizes time into short segments of duration $\delta t$ and alternates: (i) free evolution under Eq.~\eqref{eq:lindblad_metrology} for $\delta t$, and (ii) a recovery operation $\mathcal{R}$ (possibly involving ancillas, syndrome measurements, and feedback) designed to undo the effect of the noise accumulated during that segment \cite{kessler2014quantum,Arrad2014,dur2014improved,zhou2018achieving}. In the idealized limit where recovery operations can be applied arbitrarily fast and accurately compared to the probe dynamics, this ``sense--correct--sense--correct'' strategy can suppress noise to leading order in $\delta t$ while allowing coherent signal accumulation for a total time $T=M\,\delta t$ for $M$ channel uses.

\paragraph{Error-corrected Heisenberg scaling conditions.} Let $\mathcal{S}$ denote the complex linear span of all jump operators and their products
\begin{equation}
    \mathcal{S} = \text{span}\{\, I,\; L_k, L_k^{\dagger},\; L_k^{\dagger}L_j ,\,\forall j,k \,\}.
\end{equation}
Then the following is a necessary and sufficient condition for achieving Heisenberg scaling~\cite{zhou2018achieving,demkowicz2017adaptive}:
\newline\noindent\textbf{HNLS condition:} \emph{Heisenberg scaling in time ($\mathcal{J}_\text{S} \propto T^2\propto M^2$) can be restored using quantum error correction
if and only if the signal Hamiltonian $h_0$ has a nonzero component orthogonal to the Lindblad span,}
\begin{equation}
    h_0 \notin \mathcal{S}.
\end{equation}
If $h_0 \in \mathcal{S}$, then even with arbitrarily fast and perfect QEC,
the estimation precision is limited to the standard quantum limit ($\mathcal{J}_\text{S} \propto T\propto M$).
If $h_0 \notin \mathcal{S}$, an appropriate QEC code can protect against the noise while preserving the signal Hamiltonian,
enabling Heisenberg scaling. The error correcting code can be constructed explicitly given the noise model~\cite{zhou2018achieving}.

Assume we have a code $\mathcal{C}$ (potentially including ancillas), with $\Pi_\mathcal{C}$ the projector onto the codespace. A convenient sufficient condition for removing the Lindblad noise to leading order while keeping the encoded state inside $\mathcal{C}$ is the (continuous-time) QEC condition
\begin{equation}\label{eq:lindblad_QEC_conditions}
  \Pi_\mathcal{C} L_k \Pi_\mathcal{C} = \lambda_k \Pi_\mathcal{C},
  \qquad
 \Pi_\mathcal{C} L_k^\dagger L_\ell \Pi_\mathcal{C} = \mu_{k\ell}\Pi_\mathcal{C}
  \qquad \text{for all } k,\ell,
\end{equation}
for some complex scalars $\{\lambda_k\}$ and Hermitian matrix $\{\mu_{k\ell}\}$. Eq.~\eqref{eq:lindblad_QEC_conditions} is a natural analog of the Knill--Laflamme conditions [Eq.~\eqref{eq:KL}]: it ensures that, to first order in time, a suitable recovery operation can return the state to the codespace without disturbing the logical degrees of freedom \cite{zhou2018achieving}.

The signal Hamiltonian induces an effective logical generator
\begin{equation}\label{eq:logical_generator}
  h_{0,\mathrm{L}} \;=\; \Pi_\mathcal{C} h_0 \Pi_\mathcal{C} \;-\; \frac{\text{Tr}(\Pi_\mathcal{C} h_0 \Pi_\mathcal{C})}{\text{Tr}(\Pi_\mathcal{C})}\,\Pi_\mathcal{C},
\end{equation}
i.e., the traceless component of $\Pi_\mathcal{C} h_0 \Pi_\mathcal{C}$ on the codespace. The protocol is useful for metrology precisely when $h_{0,\mathrm{L}}\neq 0$, equivalently when $\Pi_\mathcal{C} h_0 \Pi_\mathcal{C}$ is not proportional to $\Pi_\mathcal{C}$. Intuitively, if $\Pi_\mathcal{C} h_0 \Pi_\mathcal{C} \propto \Pi_\mathcal{C}$, then the code hides the signal: all states in $\mathcal{C}$ acquire the same phase and no information about $\phi$ can be extracted.


\paragraph{Precision scaling under ideal QEC.}
Under the idealized assumptions above, repeated application of the recovery yields an effective logical evolution approximately generated by $h_{0,\mathrm{L}}$ over time $T$,
\begin{equation}
  \rho_{\mathrm{L}}(T) \approx e^{-i\phi T h_{0,\mathrm{L}}}\,\rho_{\mathrm{L}}(0)\,e^{+i\phi T h_{0,\mathrm{L}}}.
\end{equation}
For a pure logical initial state $\ket{\psi_{\mathrm{L}}}$, the QFI then takes the familiar unitary form
\begin{equation}\label{eq:QFI_unitary_logical}
\mathcal{J}_\text{S}\;=\;4T^2\,\text{Var}(h_{0,\mathrm{L}}),
\end{equation}
so the estimation error can scale as $\sigma_\phi = O(1/T)\propto O(1/M)$, i.e., Heisenberg scaling in interrogation time or number of channel uses.

\paragraph{SQL scaling.}
Finally we note that the optimal attainable precision when HNLS is violated has also been studied: Ref.~\cite{zhou2020optimal} showed that approximate error correction codes can remove some noise terms. More recently, Ref.~\cite{yin2025small} presented a family of quantum states that are optimal in this setting.


\subsection{Fault tolerance}
\label{subsecFT}
The discussion above treats QEC in an ideal setting: syndrome measurements are assumed to be accurate, ancillas can be prepared and reset reliably, and the recovery operation can be executed without introducing additional faults. These assumptions are appropriate for examining fundamental limits in quantum sensing, but they are also precisely where the gap between theory and experiment becomes most acute. In reality, every component of a quantum-error-corrected sensing protocol can fail: state preparation, entangling gates for syndrome extraction, idle periods, and the final measurement. The theory of fault tolerance addresses this problem by showing how to build protocols whose logical failure probability can be made arbitrarily small even when the underlying hardware is noisy, provided the physical error rate is below a nonzero threshold \cite{shor1996fault,AharonovBenOr1999,TerhalBurkard2004,Fowler2012}.

A standard starting point is a circuit-level noise model in which each location---a gate, measurement, state preparation, or idling step---fails independently with probability $p$, producing a stochastic Pauli error. A fault-tolerant construction has two particularly important features. First, it is error-local: a single physical fault should not spread into high-weight errors. This is enforced by special-purpose gadgets for syndrome extraction (often using verified ancillas or repeated measurements) and by careful compilation of logical operations. Second, it is scalable: by concatenating codes or increasing the size of a topological code, one can suppress the logical error rate superpolynomially (often exponentially) in the overhead. The threshold theorem then states that if $p<p_{\mathrm{th}}$ for some architecture-dependent threshold $p_{\mathrm{th}}$, arbitrarily long circuits can be executed reliably~\cite{AharonovBenOr1999,Fowler2012}.

A central concept in ensuring that errors in one part of the circuit do not spread to the rest of the circuit is that of transversal gates. A transversal logical gate is one that can be implemented on logical qubits by independent physical gates. Related to this, a key result in fault-tolerant quantum computing is the Eastin--Knill theorem \cite{EastinKnill2009}. This shows that no finite-dimensional quantum error-correcting code allows a universal transversal logical gate set, particularly forbidding a continuous one-parameter group of logical gates transversally. 

\subsubsection{Fault-tolerant quantum sensing}
The main motivations for bridging fault tolerance and quantum sensing are highlighted above---error-corrected quantum sensing, as in Ref.~\cite{zhou2018achieving}, relies on arbitrarily fast, noiseless operations, ideal state preparation, and measurement, and a perfectly known noise model. Removing some or all of these assumptions is central to making quantum error-corrected sensing applicable to realistic settings. Furthermore, quantum sensing often relies on complex probe states, which can require deep quantum circuits to prepare, e.g.~GHZ-type states. A single error at the start of such circuits can remove the Heisenberg scaling asymptotically.


For quantum sensing, importing fault tolerance suffers one key difficulty. The signal itself is typically a continuous-time Hamiltonian rotation. However, the Eastin--Knill theorem constrains the existence of codes with transversal implementations of continuous symmetry groups, highlighting a potential obstruction to realizing a signal Hamiltonian as a fully fault-tolerant logical operator~\cite{EastinKnill2009}. Specifically, the commonly considered Hamiltonian $H=\phi\sum_i Z_i/2$ involves a continuous $Z$ rotation implemented transversally across all qubits. From the Eastin--Knill theorem, we know that such an operation cannot correspond to a continuous logical operation for any finite dimensional code correcting arbitrary local errors.




Despite these challenges, recent work has begun to consider fault-tolerant quantum sensing, with different studies assigning different meanings to the term~\footnote{Error correction codes have also been used to estimate certain parameters of noise models~\cite{wagner2021optimal,wagner2022pauli}.}. Ref.~\cite{kapourniotis2019fault} was the first paper studying fault-tolerant quantum metrology. In this work, a threshold was defined for quantum metrology as \textquote{the strength of noise below which the estimator for the parameter converges.} However, this definition is implicitly dependent on the chosen sensing protocol~\cite{rudolph2003quantum}. While this protocol demonstrates an advantage over the corresponding non-fault-tolerant protocol and is robust against arbitrary single-qubit errors, the protocol itself is not necessarily optimal as only a finite number of bits of $\phi$ can be estimated. Another proposal for fault-tolerant sensing was put forward in Ref.~\cite{mezzadri2025dephasing}. Focusing on spin qudits, Ref.~\cite{mezzadri2025dephasing} showed how to construct a code that ensures a specific transversal rotation is also logical under a specific dephasing noise model. During the preparation of this manuscript, we became aware of another manuscript investigating fault-tolerant sensing. Ref.~\cite{sahu2026achieving} introduced a protocol which restores Heisenberg scaling and is tolerant to faults at every location in the circuit, but only for a restricted class of errors, in agreement with Ref.~\cite{zhou2018achieving}. This is similar to our proposal in Sec.~\ref{sec:FTlimitednoise} as will be elaborated on later. Restricting to a limited class of errors is the main difference between our work (also Ref.~\cite{sahu2026achieving}) and that of Ref.~\cite{kapourniotis2019fault}.

The different nature and aims of these works illustrate the challenges in identifying an operationally useful definition of fault-tolerant quantum sensing. For example, the ability to correct arbitrary single-qubit errors is incompatible with achieving Heisenberg scaling for estimating $\phi$ in Eq.~\eqref{eq:Hamiltonian}, from the HNLS condition~\cite{zhou2018achieving}. Hence, these two desirable properties cannot be achieved simultaneously. In this work, we use the term ``fault-tolerant quantum sensing'' to mean a sensing protocol capable of handling (a restricted class of) errors in every operation involved in the protocol. We note that there are many topics that are parallel to this research direction, such as sensing using existing error correcting codes~\cite{ott2026rare,antu2025stabilizer}.

\section{Basic results for Fault-Tolerant quantum sensing}
\label{sec:basicFT}
We begin by presenting some basic results regarding fault-tolerant sensing using error correction codes capable of correcting arbitrary single-qubit errors. Sec.~\ref{subsecHSEC} discusses when Heisenberg scaling is compatible with different error correcting codes, and Sec.~\ref{subsec:nonlogicalRot} discusses the effect of physical rotations not corresponding to logical rotations.

\subsection{Heisenberg scaling with error correcting codes}
\label{subsecHSEC}
We first present some results that follow trivially from Ref.~\cite{zhou2018achieving}. Nevertheless we state them here with the interpretation as results about a specific code, rather than a channel. This is primarily to provide context for the results in the subsequent section.

\subsubsection{Maximum-distance code for Heisenberg scaling in qubit number}
We are interested in estimating a transversal rotation about the $Z$ axis, corresponding to the Hamiltonian in Eq.~\eqref{eq:Hamiltonian}. For this problem it is well known that a probe state $\ket{\psi}$ achieves a QFI of
\begin{equation}
\begin{split}
        \mathcal{J}_\text{S}&=4(\langle h_0^2\rangle-\langle h_0\rangle^2)\\
        &=\sum_{i,j}\langle Z_iZ_j\rangle-\left ( \sum_{i}\langle Z_i\rangle \right )^2\\
    &=N+\sum_{i\neq j}\langle Z_iZ_j\rangle-\left(\sum_i\langle Z_i\rangle\right)^2
\end{split}
\end{equation}
where we have used the fact that $Z_iZ_i=I_2$.
Next note that, for any quantum code capable of detecting arbitrary errors on two qubits (i.e., all weight-2 Pauli errors have non-trivial syndromes), the $\langle Z_iZ_j\rangle$ term is 0. We therefore have the following corollary.
\begin{corollary}[Fault tolerance and Heisenberg Scaling]
\label{lemmaFTHS}
Any error correcting code where all weight-2 Pauli errors have non-trivial syndromes, cannot achieve Heisenberg scaling in qubit number for estimating a transversal rotation about the $Z$-axis. Specifically, $\mathcal{J}_\text{S}\leq N$, i.e.\ at best the SQL is achieved.
\end{corollary}
We stress the importance of being able to detect arbitrary two qubit errors. This excludes distance 3 codes where some $Z_iZ_j$ ($i\neq j$) can act trivially, such as the Shor code. Nevertheless, this result, although trivial, should serve as a useful guide for determining whether a given code allows Heisenberg scaling for estimating a transversal $Z$ rotation.


\subsubsection{No Heisenberg scaling in time for correcting arbitrary errors}
From Ref.~\cite{zhou2018achieving} (see also Sec.~\ref{subsecEC}), we know that necessary and sufficient conditions for achieving Heisenberg scaling are that the HNLS condition is satisfied. When this is the case, there exists a code that satisfies
\begin{equation}
    \begin{split}
        \Pi_CL_k\Pi_C&=\lambda_k\Pi_C,\;\forall k\;,\\
        \Pi_CL_k^\dagger L_j\Pi_C&=\mu_{kj}\Pi_C,\;\forall k,j\;,\\
        \Pi_C h_0\Pi_C&\neq \alpha\Pi_C\;,
    \end{split}
\end{equation}
where $\alpha$ is some constant. Trivially, if the jump operators span the whole Hilbert space,  HNLS cannot be satisfied. Consider now a code that is in principle capable of correcting errors in the same direction as the signal, i.e., for estimating a rotation about the $Z$-axis, $\Pi_CZ_i\Pi_C=\lambda_i\Pi_C$. Then it is clear that the third line in the above equation also cannot be satisfied. Hence, any error correcting code capable of correcting all errors in the same direction as the signal, cannot achieve Heisenberg scaling in signal accumulation time.  

We stress that this follows trivially from the HNLS condition~\cite{zhou2018achieving}, but for the purposes of this manuscript, we wish the reader to interpret this as a property of a code, rather than a channel.


\subsection{Effect of non-logical rotations}
\label{subsec:nonlogicalRot}
One of the main difficulties with fault-tolerant sensing, as highlighted in Ref.~\cite{kapourniotis2019fault}, is that transversal rotations will not, in general, correspond to logical rotations owing to the Eastin--Knill theorem. Again, we consider evolution under the Hamiltonian given in Eq.~\eqref{eq:Hamiltonian}. At $\phi=0$, the physical and logical rotations are equal (as this corresponds to the identity operation). However, as $\phi$ deviates from 0, the physical rotation will not correspond to a logical rotation. We can expand the rotation operator acting on the state as
\begin{equation}
\begin{split}
      \mathrm{e}^{-i\phi th_0}&=I_D+\sum_{k=1}^{\infty} \frac{1}{k!}(-i\phi th_0)^k\\&=I_D+\sum_{k=1}^{d-1} \frac{1}{k!}(-i\phi th_0)^k+\sum_{k=d}^{\infty} \frac{1}{k!}(-i\phi th_0)^k\;,
\end{split}
\end{equation}
where $d$ is the code distance and $D$ is the dimension of the Hilbert space. Below, we examine the effect of error detection on the encoded signal.
 
\subsubsection{Error detection}

For simplicity, we consider here error detection rather than error correction, i.e.~we assume the experimenter discards any runs where an error is observed. We can model this as projecting back onto the code space (setting $t=Mdt$). The unnormalized postselected state is
\begin{align}
\ket{\widetilde\psi_{\mathrm{L,fin}}}
&=
\Pi_C e^{-iM \phi dt h_0}\ket{\psi_L} \nonumber\\
&=
\sum_{k=0}^{\infty}
\frac{(-i M \phi dt)^k}{k!}
\Pi_C h_0^k\Pi_C\ket{\psi_L}.
\end{align}
Suppose that every operator appearing in $h_0^k$ for $k<d$ is
detectable by the code. The error-detection conditions then imply
\begin{equation}
        \Pi_C h_0^k\Pi_C=c_k\Pi_C,
    \qquad k=0,\ldots,d-1\;,
\end{equation}
for some constants $c_k$. These terms do not vanish; rather, they act
as scalar multiples of the identity within the code space and hence
contribute only an overall amplitude and phase to the postselected
state.

Writing $\Pi_C h_0^d\Pi_C=c_d\Pi_C+K_d$, where $K_d$ is the first component that acts nontrivially within the logical code space, we obtain
\begin{equation}
\begin{split}
        \ket{\widetilde\psi_{\mathrm{L,fin}}}
=&\alpha_{d}(M \phi dt)\ket{\psi_L}+\frac{(-iM \phi dt)^d}{d!}
K_d\ket{\psi_L}\\&+O((M \phi dt)^{d+1}),
\end{split}
\end{equation}
where $\alpha_d(M \phi dt)$ is a scalar. After normalizing the
postselected state and discarding an irrelevant global phase, the
first nontrivial encoded evolution is therefore of order $O\!\left((\phi t)^d\right)$.
Evidently, error detection with higher distance codes exponentially suppresses any encoded signal, making it harder to find regimes where such codes can offer an advantage in sensing. Note that error correction leads to a similar suppression of the signal.


\subsubsection{Comparison to existing work}
One might imagine that the suppression of the signal, highlighted by the noiseless case studied above, would be difficult to overcome. However, Ref.~\cite{ott2026rare} has recently suggested that error correcting codes can be useful for sensing transversal operations. Specifically, Ref.~\cite{ott2026rare} works in a regime where the signal to noise ratio is very low (rare event sensing) such that the suppression of the signal discussed above is compensated for by the corresponding decrease in noise, to get an overall improvement in sensing performance.

\section{Fault-tolerant sensing for restricted noise model}
\label{sec:FTlimitednoise}
We now present our main result, examining a setting inspired by the HNLS condition of Ref.~\cite{zhou2018achieving}, and the above discussion, with the goal of removing the assumption of perfect control operations. It should be noted that Ref.~\cite{zhou2018achieving} does allow for noise acting on the ancilla states during the evolution. (The requirement of ancillas can be removed in certain cases~\cite{layden2019ancilla,zhou2024achieving}.) However, the quantum error correcting code developed in Ref.~\cite{zhou2018achieving} does not automatically specify a quantum circuit to prepare the quantum state of interest, extract the syndrome measurement, implement the necessary correction operations or perform a measurement at the end. As such, fault-tolerant circuits that are robust against more general noise models cannot be directly deduced (or necessarily expected) from Ref.~\cite{zhou2018achieving} in general. In accordance with the HNLS condition, fault-tolerant Heisenberg scaling cannot be expected for arbitrary noise models. Following this, Ref.~\cite{sahu2026achieving} has recently developed a fault-tolerant protocol which achieves Heisenberg scaling in qubit number even when allowing for faults at any location in the quantum circuit. Specifically, they allow $X$ (bit-flip) errors at any location in the circuit (state preparation, gate operation, idling, and measurement), and show that for estimating a rotation about the $Z$-axis, Heisenberg scaling can be achieved. Here we consider similar examples, and demonstrate that Heisenberg scaling can be achieved with sequential evolution rather than parallel as in Ref.~\cite{sahu2026achieving}. As we shall discuss later, this distinction can lead to significantly enhanced performance, especially with a restricted number of qubits.

\subsection{Channel model}
We consider a $N$-qubit independent bit-flip channel which acts on each qubit with probability $p$ before the unknown rotation (about the $Z$ axis) acts only on the first qubit. The Hamiltonian is therefore
\begin{equation}
    H_1=\phi Z_1/2\;,
\end{equation}
where the subscript 1 denotes that the rotation is only on the first qubit. The total channel acting on a state $\rho$ per interaction with the channel is then
\begin{equation}
    \mathcal{E}(\rho)=\mathrm{e}^{-iH_1}\mathcal{E}_\text{BF}(\rho)\mathrm{e}^{iH_1}\;,
\end{equation}
where $\mathcal{E}_\text{BF}(\rho)$ denotes a bit-flip error with probability $p$ on every qubit, and without loss of generality we have set the interaction time per channel use to be 1. If we considered an alternative model whereby the bit-flips occur only after the rotation, the results that follow remain qualitatively unchanged. A more general model, where bit-flip errors can occur throughout the evolution time is considered in appendix~\ref{apen:discretetime}. Note that, as we are assuming the channel implements a bit-flip with probability $p$ (rather than a continuous process as in Ref.~\cite{zhou2018achieving}), we are effectively disallowing arbitrarily fast quantum control. We assume that any ancilla qubits are subject to the same noise channel. We allow the experimenter to interact with the channel $M$ times, and they are free to reset or measure the probe state at any stage in the evolution. We set $M=M_1\nu$, such that $M_1$ is the number of channel uses experienced before the probe state is measured and reset, and the process is repeated $\nu$ times.

\begin{figure}[t]
    \centering
\includegraphics[width=0.9\columnwidth]{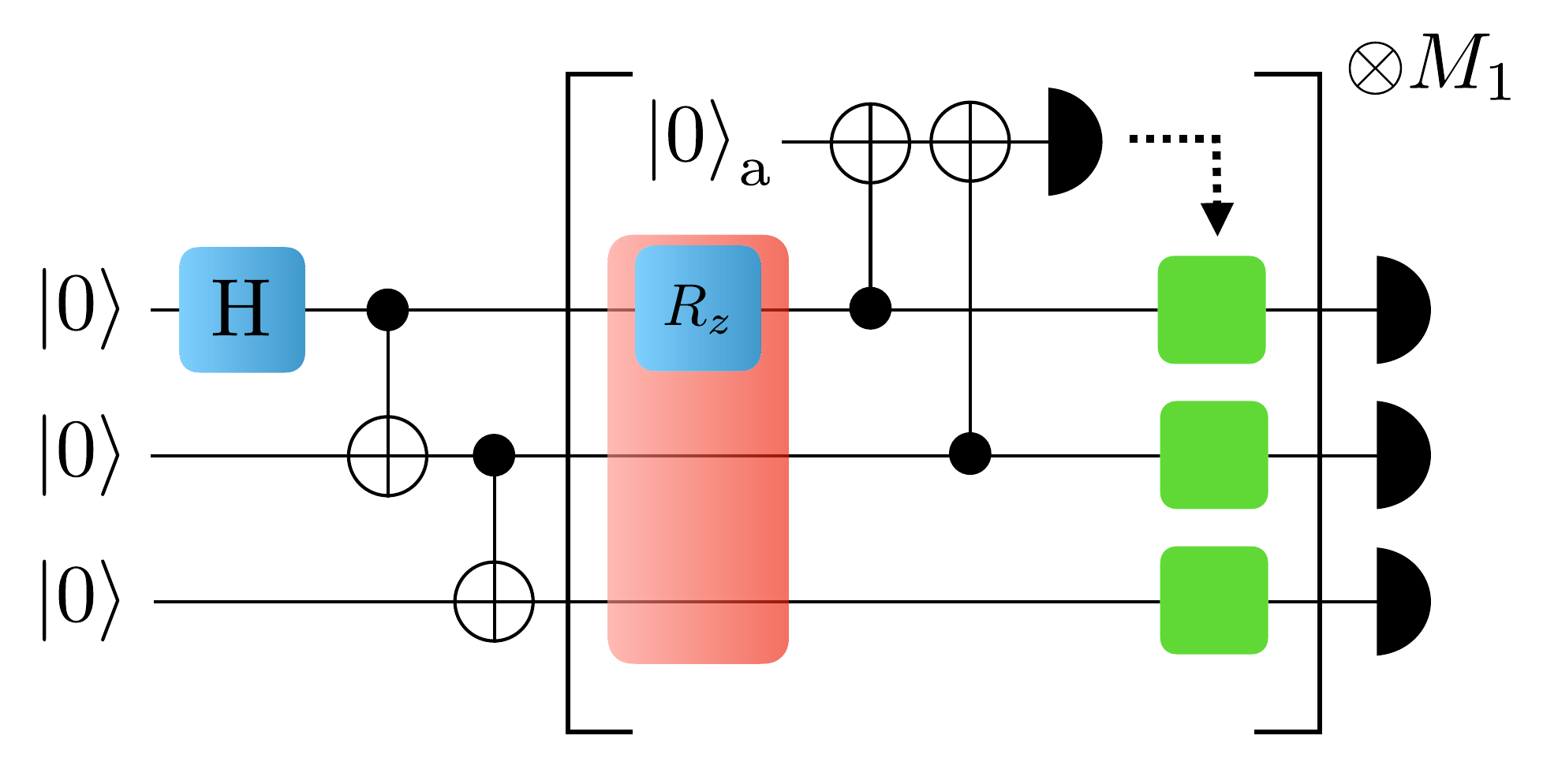}
    \caption{\textbf{Ideal error-corrected sensing with a three-qubit repetition code.} In the ideal setting, errors occur only in the quantum channel, shown as a red box and all other operations are assumed to be perfect. The green boxes denote a correction operation that depends on the syndrome measurement outcomes. For simplicity the syndrome measurement $Z_2Z_3$ is not shown.}
    \label{fig:ECideal}
\end{figure}

Under this noise model, with no ancilla qubits or error correction, Heisenberg scaling can be achieved up to a number of channel uses $M_1\propto 1/p$ (see appendix~\ref{subsec:signoisediff}). Below, we first describe the sensitivity limits using ideal error correction, i.e. without errors at every location in the circuit (standard error-corrected quantum metrology~\cite{zhou2018achieving}, see Fig.~\ref{fig:ECideal}). We then discuss a protocol that works even when this assumption is removed (i.e. fault-tolerant quantum metrology, see Fig.~\ref{fig:FTnaive}). \\

\subsection{Ideal error-corrected sensing---errors only in channel}
\label{subsec:idealsetting}
With no errors in state preparation, we can prepare an ideal GHZ state
\begin{equation}
\ket{\psi_\text{GHZ}}=\frac{\ket{0}^{\otimes N}+\ket{1}^{\otimes N}}{\sqrt{2}}\;.
\end{equation}
In the bit-flip noise model considered here, $\ket{\psi_{\mathrm{GHZ}}}$ is the logical $|+\rangle_L$ state of the $N$-qubit repetition code. Going forward, the term distance is used in the classical sense relevant to $X$-type errors: the repetition code has distance $N$ against bit flips, and hence corrects up to $t=\lfloor (N-1)/2 \rfloor$ such errors. When interacting with the quantum channel, the bit-flip channel creates the state
\begin{equation}
\label{eq:ghzallbitflip}
    \rho_{\text{GHZ}}=\sum_{k=0}^{N}p^k(1-p)^{N-k}\sum_{j=1}^{\binom{N}{k}}\rho_{\text{GHZ},k,j}\;,
\end{equation}
where $\rho_{\text{GHZ},k,j}$ represents all the different possible ways $k$ $X$ operators may have acted upon the GHZ state. The rotation then acts only on the first qubit to give
\begin{equation}
    \rho_{\text{GHZ}}(\phi)=\sum_{k=0}^{N}p^k(1-p)^{N-k}\sum_{j=1}^{\binom{N}{k}}\rho_{\text{GHZ},k,j}(\phi)\;,
\end{equation}
where $\rho_{\text{GHZ},k,j}(\phi)=\mathrm{e}^{-iH_1}\rho_{\text{GHZ},k,j}\mathrm{e}^{iH_1}$. By projecting onto spaces with $k$ bits flipped, we can correct errors on up to $t$ qubits. The projectors are denoted $P_{k,i}$, where the $k$ denotes how many bits are flipped and $i$ indexes over all possible ways these bits can be flipped. For example, 
\begin{widetext}
    \begin{equation}
    \begin{split}
        P_{0}&=(\ket{0}\bra{0})^{\otimes N}+(\ket{1}\bra{1})^{\otimes N}\;,\\
        P_{1,1}&=(\ket{1}\ket{0}^{\otimes N-1})(\bra{1}\bra{0}^{\otimes N-1})+(\ket{0}\ket{1}^{\otimes N-1})(\bra{0}\bra{1}^{\otimes N-1})\;.
    \end{split}
\end{equation}
\end{widetext}
Denoting ${t'}=t+1$, we can write the logical error probability as
\begin{equation}
    p_\text{L}=\sum_{k=t'}^{N}\binom{N}{k}p^k(1-p)^{N-k}\approx c_{t'}p^{t'}+O(p^{t'+1})\;,
\end{equation}
where $c_{t'}$ denotes all the possible ways $t'$ errors can occur. The state resulting from this error detection and correction is
\begin{equation}
    \rho_{\text{GHZ},1}(\phi)=(1-p_\text{L})\ket{\psi_\text{GHZ}(\phi_1)}\bra{\psi_\text{GHZ}(\phi_1)}+p_\text{L}\tilde{\rho}\;,
\end{equation}
where the subscript 1 in $\rho_{\text{GHZ},1}(\phi)$ denotes that this is after a single channel use, and $\tilde{\rho}$ contains all states that could have caused a logical error. The subscript 1 in $\phi_1$ denotes that the unknown phase depends on the correction operation applied. If the first bit was flipped the phase is $-\phi$, and if it was not flipped the phase is $\phi$. In general we need to keep track of all syndrome measurements to produce an unbiased estimate of $\phi$ at the end of the procedure. However, this does not affect the scaling in number of channel uses, as the effective error has been reduced from $p$ to $p_\text{L}=c_{t'}p^{t'}$. After $M_1$ uses of the channel, we get the state
\begin{align}
    \rho_{\text{GHZ},M_1}(\phi)=&(1-p_\text{L})^{M_1}\ket{\psi_\text{GHZ}(\phi_{M_1})}\bra{\psi_\text{GHZ}(\phi_{M_1})}
    \nonumber\\
\label{eq:GHZideal}
    &+(1-(1-p_\text{L})^{M_1})\tilde{\rho}\;,
\end{align}
where $\tilde{\rho}$ has been redefined appropriately, and $\phi_{M_1}$ is the error-dependent accumulated phase after $M_1$ rounds.

When extracting information about $\phi$ we need to be careful about the information contained in $\tilde{\rho}$. $\tilde{\rho}$ is a classical mixture of GHZ states with different phases, where the phase depends on the specific errors that occurred. Naively measuring the total accumulated phase without accounting for the error dependence may introduce a bias into the estimate of $\phi$, potentially eliminating any improvement. A method for constructing an unbiased estimator is described in appendix~\ref{apen:unbiasedest}. Demonstrating the existence of an unbiased estimator alone is not sufficient to show that Heisenberg scaling can be achieved.
Two issues remain---the signal cannot be suppressed too much and the variance cannot be too large. Accordingly, in appendix~\ref{apen:problemreduction}, we bound the variance of the unbiased estimator introduced in appendix~\ref{apen:unbiasedest}, showing that it is increased by a factor of at most $1/(2(1-p_\text{L})^{M_1}-1)^2\approx1/(1-2M_1p_\text{L})^2$, for $1/2<(1-p_\text{L})^{M_1}<1$, compared to the ideal case. Next consider the amplitude of the signal observed. If $N_x=N_x(M_1,p)$ bit flips occurred on the first qubit in the $M_1$ channel uses then 
\begin{equation}
\phi_{M_1}=(M_1-N_x)\phi-N_x\phi=(M_1-2N_x)\phi\;.
\end{equation} 
Note that, conditioned on successful error correction
\begin{equation}
\label{eq:expectationphiM1}
\langle \phi_{M_1}\rangle = M_1\phi(1-2p)\;,
\end{equation}
as with probability $1-p$ the phase in a given channel use was $\phi$ and with probability $p$ it was $-\phi$. Hence, the accumulated phase is reduced by a factor $(1-2p)$. Note that here we assume that $N_x\sim\mathrm{Binomial}(M_1,p)$, implying that $\phi_{M_1}$ is a random variable, contributing an extra dephasing factor. Throughout the remainder of this manuscript we therefore require $|\phi|\lesssim1/M_1$, which ensures this effect is negligible (see appendix~\ref{apen:errordet}).


Using the estimation procedure outlined in appendix~\ref{apen:unbiasedest}, we arrive at a Fisher information for estimating $\phi$ of (to leading order in $M_1$)
\begin{equation}
\label{eq:idealHS}
   \mathcal{J}\geq M_1^2(2(1-p_\text{L})^{M_1}-1)^2(1-2p)^2\;.
\end{equation}
This CFI is maximized when $M_1\approx-0.315/\log(1-p_\text{L})\approx 0.315/p_\text{L}$, recovering the discrete time analog of Heisenberg limited scaling up until a certain time\footnote{Starting from Eq.~\eqref{eq:GHZideal} if one simply ignores $\tilde{\rho}$ (for example by claiming $M_1\ll1/p_\text{L}$) then one could arrive at a very similar expression that contains all the essential physics.}. However, the corresponding variance must account for the total number of channel uses $M=M_1\nu$, and is optimized when $M_1\approx-0.2/\log(1-p_\text{L})\approx a/p_\text{L}$ for $a=0.2$. The variance at this optimal point is given by
\begin{equation}
\label{eq:idealHS2}
\begin{split}
    \sigma_\phi^2&\approx\frac{1}{\nu \mathcal{J}}\leq\frac{60p_\text{L}^2}{\nu(1-2p)^2}\;, 
\end{split}
\end{equation}
 which gives Heisenberg scaling in $M_1$ up to this optimal value. The bound derived in Eq.~\eqref{eq:idealHS2} is only valid for $1/2<(1-p_\text{L})^{M_1}<1$. The reason for this restricted range of validity is because we are using a worst-case bound, derived in appendix~\ref{apenworstcase}. In the worst case, for $M_1>-\log(2)/\log(1-p_\text{L})$, the QFI can vanish (and therefore the CFI will vanish).

Before proceeding to errors in state preparation and measurement, we recall the bound on the QFI when using a single-qubit state, $\ket{+}$, discussed in appendix~\ref{apenA4}. From Eq.~\eqref{eq:QFIsinglequbitXnoise}, we know that Heisenberg scaling can be recovered up until times (equivalently number of channel uses) proportional to $1/p$. Hence the power of ancilla states and entanglement is clear in the case where noise is present only in the channel. Below we demonstrate that the utility of ancilla states and entanglement extends to the case where noise is present everywhere.

\subsection{Circuit-level errors}
\label{subsec:circuitlevel}
\begin{figure}[t]
    \centering
\includegraphics[width=0.99\columnwidth]{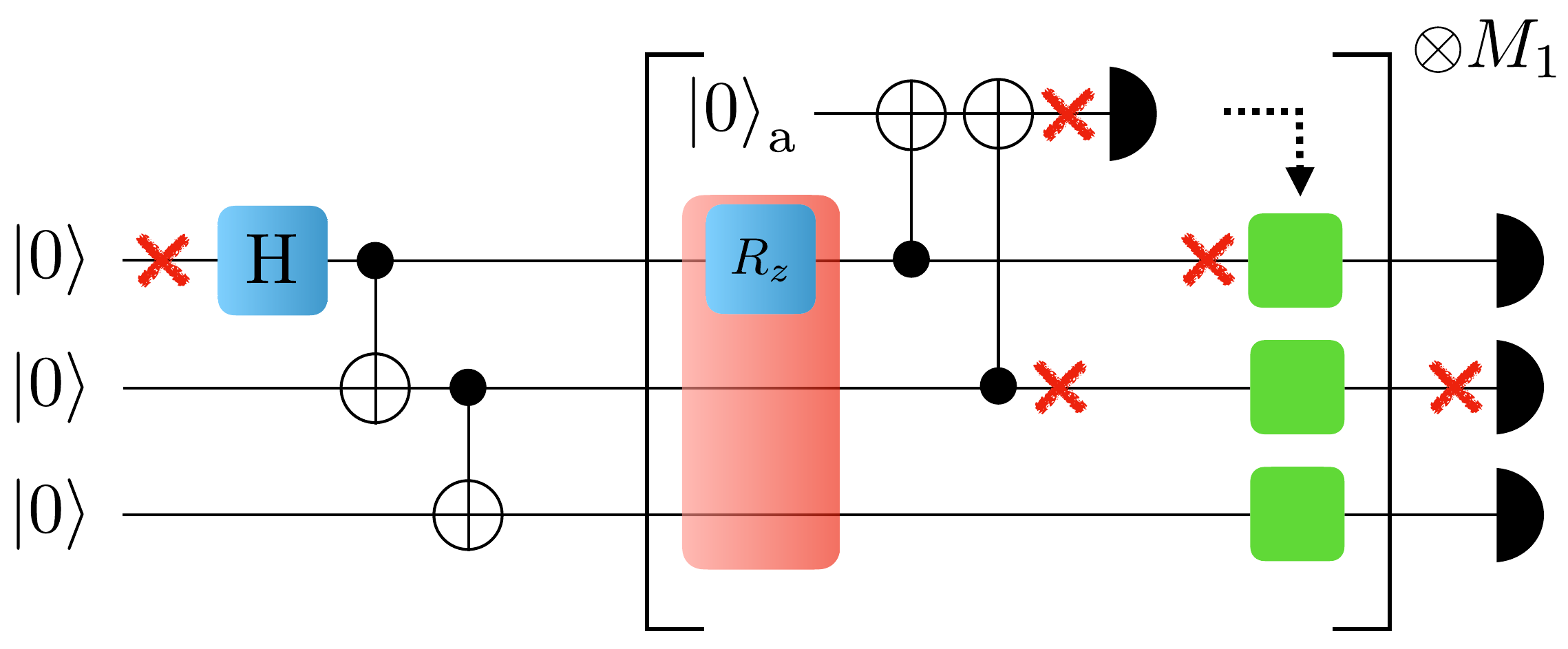}
    \caption{\textbf{Naive sensing protocol failures.} Crosses mark parts of the quantum circuit where a single physical error can cause a standard quantum-error-correction-assisted quantum metrology protocol to fail completely. Only some error locations are shown. The circuit required to implement the final measurement is not shown explicitly.}
    \label{fig:FTnaive}
\end{figure}

We now elaborate on how physical (circuit-level) errors can immediately cause the loss of Heisenberg limited scaling unless we do things fault-tolerantly. Our noise model includes the following components:
\begin{enumerate}
    \item \textbf{State preparation errors.} When preparing the $\ket{0}$ or $\ket{+}$ state, the states $\ket{1}$ or $\ket{-}$ are respectively created with probability $p$, and vice versa.
    \item \textbf{Gate and idling errors.} After every gate operation (one-qubit gate, two-qubit gate, or idling), every qubit suffers independent bit flips with probability $p$.
    \item \textbf{Measurement errors.} We only allow measurements in the computational basis, and measurement results are flipped with probability $p$.
\end{enumerate}
The more general case with unequal error rates is discussed in appendix~\ref{apen:faultolgeneral}.

We begin by discussing errors in state preparation. The red crosses in Fig.~\ref{fig:FTnaive} represent some locations where a single physical error would cause a logical error. The GHZ state needed for error-corrected sensing is naively prepared using a single Hadamard gate followed by an array of CNOT gates. Evidently an error before the Hadamard gate, which occurs with probability $p$ will propagate to a logical error. This is true even assuming perfect CNOT gates. Therefore the $\sigma_\phi\propto p_L\propto p^{t'}$ scaling from above will immediately be compromised. This is true when even more efficient circuits for GHZ generation are used~\cite{cruz2019efficient,chen2023short}.

Another important stage of a quantum error correcting metrology protocol is the syndrome extraction and corresponding correction. This syndrome extraction is normally done using controlled operations to an ancilla qubit. As shown in Fig.~\ref{fig:FTnaive}, there are several stages where this operation may go wrong.

Finally, we examine errors in the measurement. For simplicity assume that everything prior to the measurement is perfect, i.e., the state right before measuring is $\ket{\psi_\text{GHZ}(\phi)}$. Measuring in the computational basis reveals no information about $\phi$. There are two main ways we can consider extracting the information. First, we could act a Hadamard gate on every qubit and then distinguish the channels by even and odd parity measurement results. However, observe that if one Hadamard fails, or one detector flips its measurement outcome, the observed parity is flipped. The overall error arising from this effect will be a function of $p$ and $N$, but is independent of $M_1$. Hence, measurement errors do not change the scaling of the MSE with $M_1$. Nevertheless, the CFI decays exponentially in $N$, making this measurement undesirable. The second method of extracting the information would be to undo the GHZ state creation, to generate a single-qubit state that contains all the information about $\phi$. The impact of this second measurement method on sensing performance will be examined in detail later (see appendix~\ref{apen:FTmeas}), however the CFI is independent of $N$, motivating the use of this second measurement method.

\subsection{Fault-tolerant sensing---bit-flip errors only}

\begin{figure}[t]
    \centering
\includegraphics[width=0.99\columnwidth]{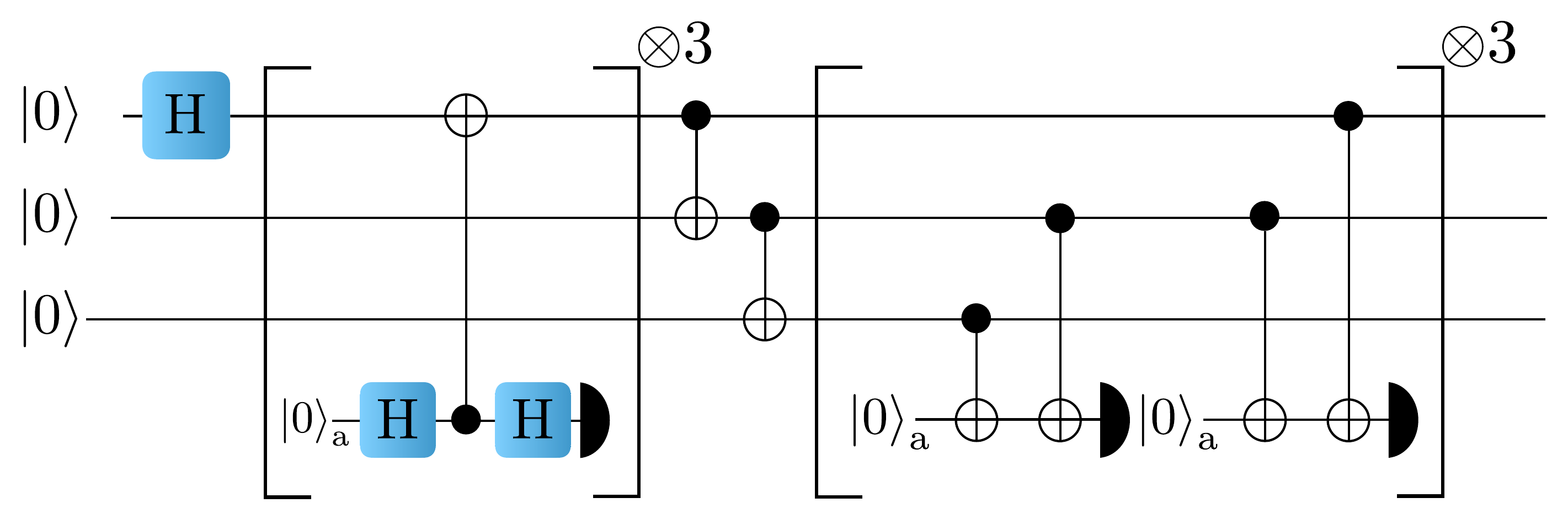}
    \caption{\textbf{Fault-tolerant state preparation.} Syndrome measurements are used to detect $X$ and $Z$ errors. The corresponding correction operations are not shown for simplicity. See appendix~\ref{apen:stateprep} for details.}
    \label{fig:FTstateprep}
\end{figure}

Given that single physical errors can create logical errors in the standard GHZ sensing protocol, this appears to be a natural regime for fault-tolerant sensing. For simplicity, we first describe the procedure with a 3-qubit GHZ state. The aim is therefore to recover Heisenberg limited scaling for a number of channel repetitions which scales as $O(1/p)<M_1<O(1/p^2)$. The general case will be considered later. 

Under our bit-flip-only circuit-level noise model, the only state-preparation fault that can produce the wrong GHZ phase ($\ket{000}-\ket{111}$) occurs at the initial $\ket{+}$ state preparation step. Under our noise model, this is the only location (prior to the final measurement) at which phase-like errors can occur. The reason for this is that $HXH=Z$, so an $X$ fault preceding the initial Hadamard gate creates a phase-like error. In contrast, faults at every subsequent location remain as bit-flip errors. We therefore first verify whether $\ket{+}$ was prepared correctly by measuring $X_1$ using an ancilla qubit (see Fig.~\ref{fig:FTstateprep} and appendix~\ref{apen:stateprep} for details). Repeating this measurement three times ensures the error in measurement is $O(p^2)$. Applying the appropriate correction ensures that the qubit is in the $\ket{+}$ state with probability $1-O(p^2)$. As this is the only point where phase-like errors can occur, this suppresses phase errors to $O(p^2)$.

We then proceed with the GHZ fanout, preparing a GHZ state with infidelity $O(p)$. Using an ancilla qubit, we then measure the stabilizers $Z_iZ_{i+1}$ three times for every $i$, see Fig.~\ref{fig:FTstateprep}. For simplicity, here we have assumed that the probability of a measurement bit-flip is also $p$. In appendix~\ref{apensyndromerepetitions}, we provide more detail on how many syndrome repetitions are needed in general. The fault-tolerant syndrome extraction and correction described below can be used at this stage if deterministic state preparation is required. This ensures that we prepare a 3-qubit GHZ with errors scaling as $\tilde{p}_{\text{L},\text{p},3}=\tilde{c}_{2,\text{p}}p^2+O(p^3)$, where the subscript p denotes state preparation and the 3 denotes that this is an $N=3$ qubit repetition code. (Alternatively, one could assume that, whenever the syndrome measurements reveal that an error occurred, we discard the GHZ state and repeat the state preparation procedure, which gives $\tilde{p}_{\text{L},\text{p},3}=O(p^3)$.) This is described in more detail in appendix~\ref{apen:stateprep}. The constant $\tilde{c}_{2,\text{p}}$ depends on the number of rounds of repetition of the syndrome extraction.

\begin{figure}[t]
    \centering
\includegraphics[width=0.9\columnwidth]{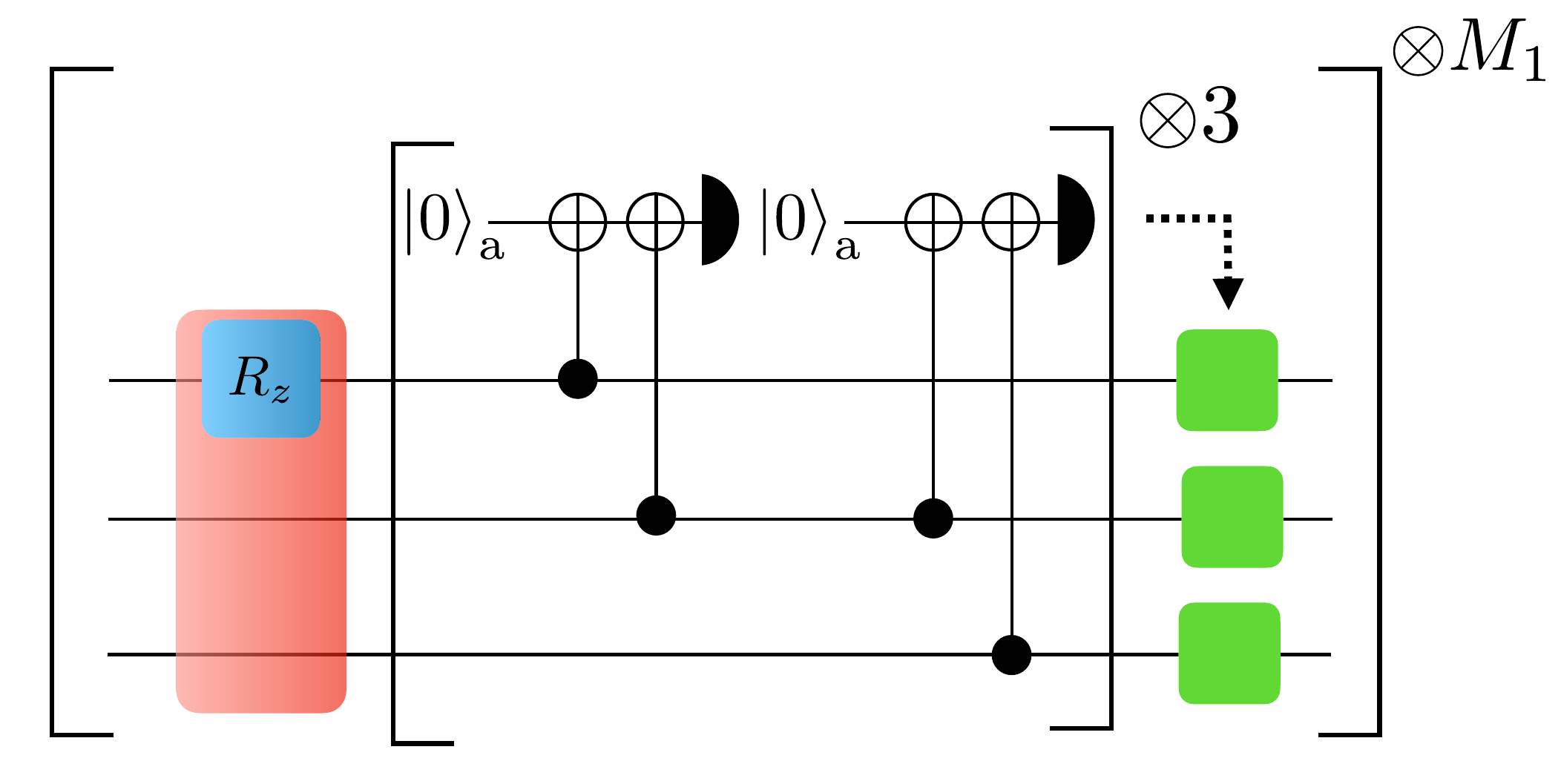}
    \caption{\textbf{Fault-tolerant syndrome extraction and error correction.} Syndrome measurements are used to detect $X$ errors that occur either during the natural evolution of the quantum state or during the correction process. The dashed arrow denotes a classical decoding process informing the appropriate correction operation shown in green. More detail on the decoding process is provided in appendix~\ref{apen:errordet}.}
    \label{fig:FTsyndromeextract}
\end{figure}

This approximate GHZ state then interacts with the quantum channel which first introduces $X$ errors on every qubit with probability $p$ followed by a rotation of the first qubit, shown in Fig.~\ref{fig:FTsyndromeextract}. After each channel use, errors are detected in a fault-tolerant manner, by coupling pairs of qubits to an ancilla qubit, measuring the ancilla, and repeating this syndrome extraction multiple times. Once an error has been identified, we implement the appropriate correction (an $X$ operation in the appropriate place). This ensures the errors are suppressed to $O(p^2)$, see appendix~\ref{apen:errordet}. As the errors are suppressed beyond the standard setup, we can coherently accumulate a phase for longer, i.e.~the number of repetitions of the channel $M_1$ can be larger. We denote the logical error probability per round as
\begin{equation}
\tilde{p}_{\text{L},3}= \tilde{c}_{2,\text{s}}p^{2}+O(p^3),
\end{equation}
where $\tilde{c}_{i,\text{s}}$ denotes all the possible ways logical errors can happen with probability that scales as $p^i$ during the interaction with the sensing channel and subsequent syndrome extraction.


\begin{figure}[t]
    \centering
\includegraphics[width=0.45\columnwidth]{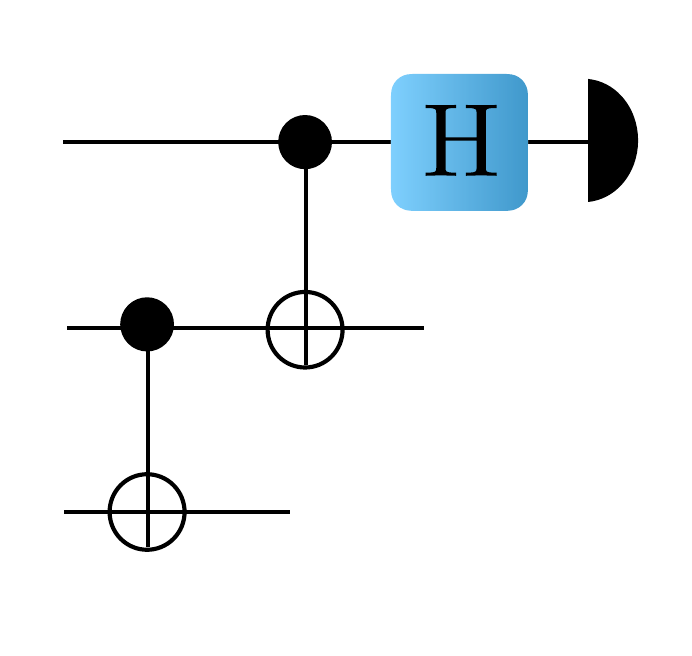}
    \caption{\textbf{Conventional GHZ measurement.} To extract information about $\phi$, we uncompute the GHZ state and measure a single qubit in the computational basis. }
    \label{fig:FTmeas}
\end{figure}

At this stage, an additional fault mechanism becomes relevant for fault-tolerant sensing beyond the logical error probability $\tilde{p}_{\text{L},3}$ discussed above. The logical errors discussed to this point have the effect of flipping the sign of any phase accumulated up until that point. That is after $M_i$ rounds the errors discussed thus far flip the sign of $\phi_{M_i}$. A second type of error, that occurs with probability $O(p^2)$, is when the sensing qubit suffers an error before the rotation channel and again after the first CNOT in syndrome extraction. We stress that this type of error occurs with probability $O(p^2)$ regardless of code distance. Hence this analysis, while overkill for the 3-qubit repetition code, is essential for higher distance codes. The resulting syndrome measurements in this case are indistinguishable from the case where there was no error on the sensing qubit but an error on the first ancilla qubit with probability $O(p)$ during the first syndrome measurement. Thus, there is some ambiguity as to whether the phase acquired in that round was $\pm\phi$. This effect is important for two reasons: 1) it must be accounted for when constructing an unbiased estimator, and 2) it will act as an additional decoherence term. However, as we show in appendix~\ref{apen:errordet}, neither of these effects impacts the attainable precision in the regime of interest. Putting everything together, after $M_1$ uses of the channel, we get the state
\begin{widetext}
    \begin{equation}
    \label{eq:FT3qubitpremeas}
    \rho_{\text{GHZ},M_1}(\phi)=(1-\tilde{p}_{\text{L},\text{p},3})(1-\tilde{p}_{\text{L},3})^{M_1}\ket{\psi_\text{GHZ}(\phi_{M_1})}\bra{\psi_\text{GHZ}(\phi_{M_1})}+(1-(1-\tilde{p}_{\text{L},\text{p},3})(1-\tilde{p}_{\text{L},3})^{M_1})\tilde{\rho}\;,
\end{equation}
where $\ket{\psi_\text{GHZ}(\phi_{M_1})}$ now denotes a state within a correctable error of the ideal GHZ state.

The final stage is then to measure the state. As we discuss in appendix~\ref{apen:FTmeas}, a simple measurement is sufficient for the three-qubit GHZ state, see Fig.~\ref{fig:FTmeas} for a pictorial representation. The fact that this is not a fault-tolerant measurement does not affect the overall scaling. Following the same logic as used for the ideal case (appendices~\ref{apen:unbiasedest} and \ref{apen:problemreduction}), we argue there is an unbiased estimator for estimating $\phi$ with this density matrix. In appendix~\ref{apen:FTmeas}, we then show that the effect of the non-fault-tolerant measurement is to reduce the CFI by a factor $(1-2p)^4$. Combining with the reduction in signal amplitude [Eq.~\eqref{eq:expectationphiM1}] and increase in variance [Eq.~\eqref{eq:idealHS}] from the ideal setting, we arrive at a CFI of
\begin{equation}
    \mathcal{J}\geq M_1^2(1-2p)^6(2(1-\tilde{p}_{\text{L},\text{p},3})(1-\tilde{p}_{\text{L},3})^{M_1}-1)^2\;.
\end{equation}
For $M=M_1\nu$, we can choose $M_1=a/\tilde{p}_{\text{L},3}$ with $a=0.2$ giving the following bound on the variance
\begin{equation}
\begin{split}
    \sigma_\phi^2&\approx\frac{1}{\nu \mathcal{J}}\leq\frac{60\tilde{p}_{\text{L},3}^2}{\nu(1-2p)^6}\;.
\end{split}
\end{equation}
\end{widetext}
 We now observe Heisenberg scaling for $M_1\approx1/\tilde{p}_{\text{L},3}=O(1/p^2)$.


More generally, we can consider an $N$-qubit repetition code, described in detail in appendix~\ref{apen:nqubitrepetition}. One major difference is that the number of repetitions for the syndrome extraction increases linearly with $N$ (see appendix~\ref{apensyndromerepetitions}). We denote the logical error probability in state preparation as $\tilde{p}_{\text{L},\text{p},N}$ and the logical error probability per round as
\begin{equation}
\tilde{p}_{\text{L},N}=\tilde{c}_{t',\text{s}}p^{t'}+O(p^{t'+1})\;.
\end{equation}
We note that $\tilde{c}_{t',\text{s}}$ will depend on $N$ and the exact syndrome extraction process. As shown in appendix~\ref{apen:nqubitrepetition}, we arrive at the following expression for the CFI: 
\begin{equation}
\label{eq:FTCFI}
    \mathcal{J}\geq M_1^2(1-2p)^6(2(1-\tilde{p}_{\text{L},\text{p},N})(1-\tilde{p}_{\text{L},N})^{M_1}-1)^2\;.
\end{equation}
Choosing $M_1=a/\tilde{p}_{\text{L},N}$, we get Heisenberg scaling in $M_1$. This bound applies when $2(1-\tilde{p}_{\text{L},\text{p},N})(1-\tilde{p}_{\text{L},N})^{M_1}-1>0$, or equivalently when $(1-\tilde{p}_{\text{L},\text{p},N})(1-\tilde{p}_{\text{L},N})^{M_1}>1/2$. The benefit of going to larger repetition codes is that the regime in which Heisenberg scaling is achievable is extended exponentially in the code distance, i.e. for a $N$-qubit repetition code, Heisenberg scaling can be achieved for $M_1=O(1/p^{t'})=O(1/p^{\lfloor(N-1)/2\rfloor+1})$.

\paragraph{Attainability of the Heisenberg limit.}
We comment briefly on the attainability of the Heisenberg limit under this model. Heisenberg scaling is achieved in $M_1$ for $M_1\leq a/\tilde{p}_{\text{L},N}\approx a/(\tilde{c}_{t'}p^{t'})\approx a/(\tilde{c}_{t'}p^{N/2})$, where the last approximation is valid for large $N$. Interestingly, to achieve Heisenberg scaling asymptotically in $M_1$ does not require $p\to0$: for a threshold noise level $p_\text{th}$, we require $p<p_{\text{th}}$ and $N$ to grow fast enough with $M_1$ such that the above inequality is satisfied. This threshold behaviour is observed in Fig.~\ref{fig:FTerrorplot2}. As discussed in appendix~\ref{apen:errordet}, we also require $\phi$ to satisfy $\phi^2p^2\ll O(\tilde{p}_\text{L})$.


\begin{figure}[t]
    \centering
\includegraphics[width=0.99\columnwidth]{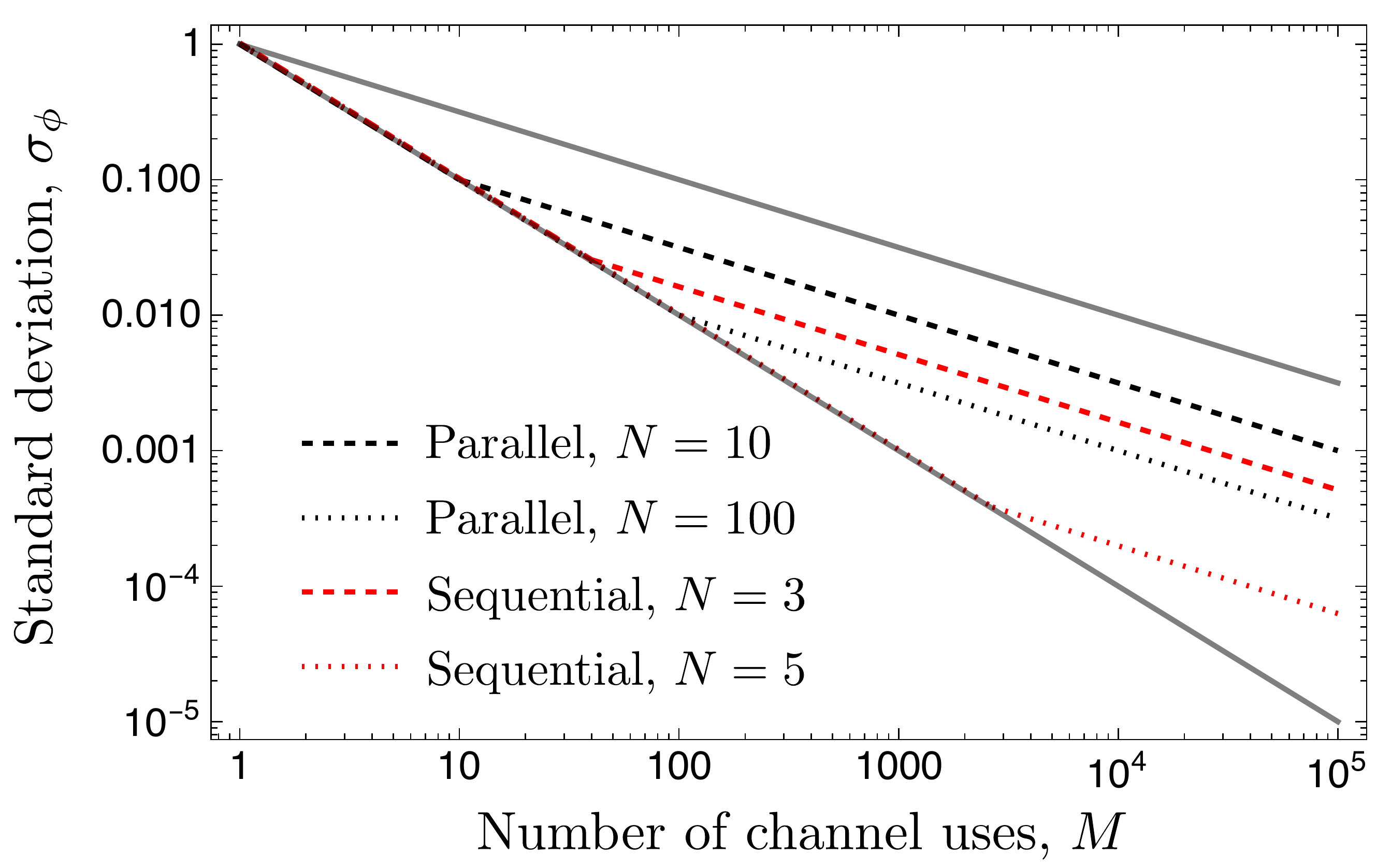}
    \caption{\textbf{Attainable error with fault-tolerant sensing.} The error with which $\phi$ can be estimated, $\sigma_\phi$, is plotted as a function of the number of channel uses. The upper and lower solid grey lines show the SQL ($1/\sqrt{M}$) and Heisenberg limits ($1/M$) respectively. For comparison we show the attainable error when performing parallel rather than sequential sensing with 10 and 100 qubits in a completely noiseless setting. For the sequential setting all gate and measurement error rates are set to be $p=2\times10^{-3}$.}
    \label{fig:FTerrorplot}
\end{figure}

\subsection{Comparison with other protocols}
\label{sec:comparison}
\begin{figure}[t]
    \centering
\includegraphics[width=0.99\columnwidth]{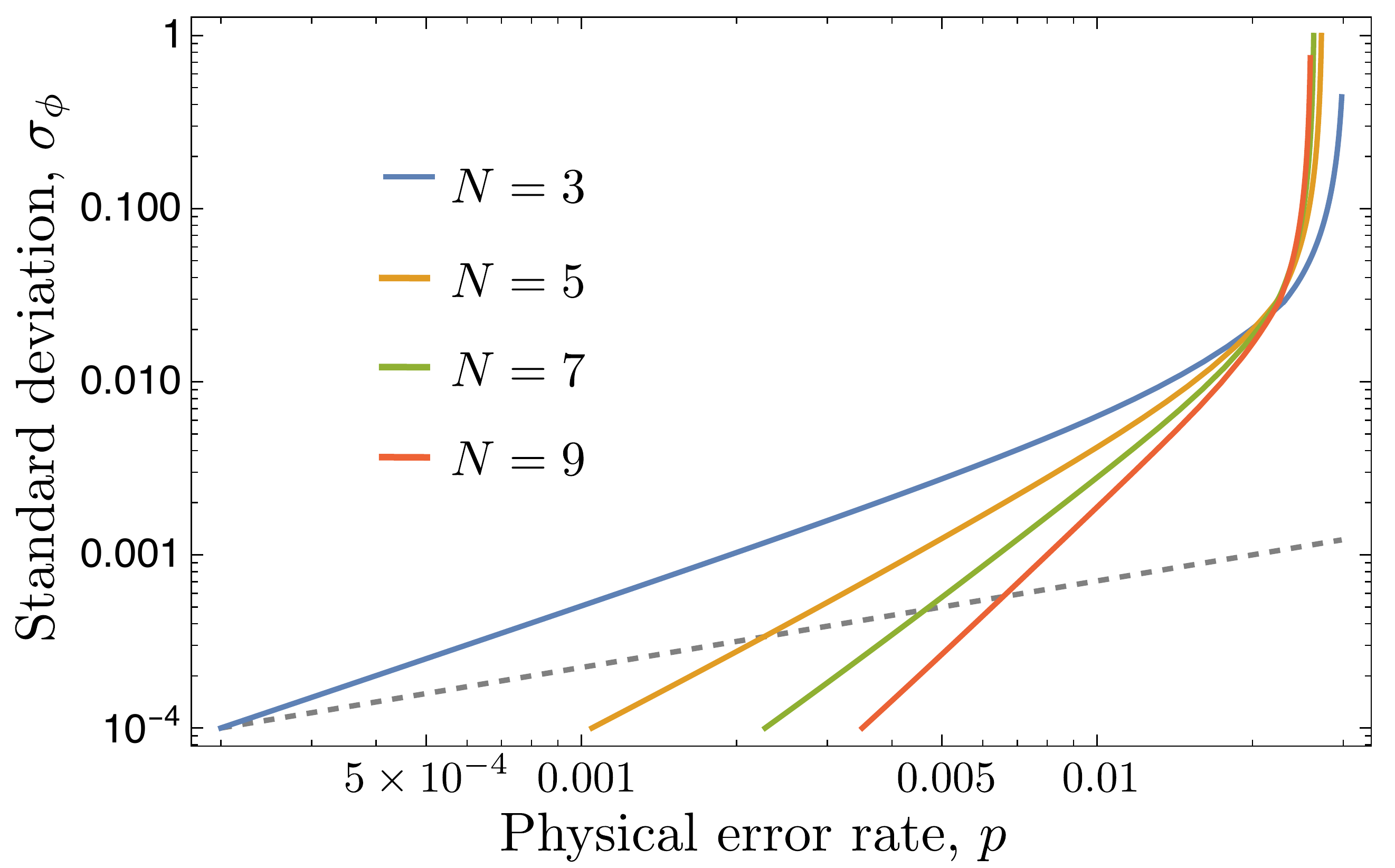}
    \caption{\textbf{Threshold in fault-tolerant quantum sensing.} The error with which $\phi$ can be estimated, $\sigma_\phi$, is plotted as a function of the physical error rate $p$. Logical error rates are obtained using STIM~\cite{gidney2021stim}. $M_1$ is set to $0.2/\tilde{p}_{\text{L},N}$ for all plots. The dashed grey line shows the optimal standard deviation ($1/(M_1\sqrt{\nu})$ with $M_1=1/p$) without entanglement. The total number of channel uses is $M=\nu M_1=20000$.}
    \label{fig:FTerrorplot2}
\end{figure}

We first compare the CFI of our fault-tolerant protocol [Eq.~\eqref{eq:FTCFI}] to the ideal quantum error-corrected sensing case [Eq.~\eqref{eq:idealHS}]. There are two main differences. The first is that, due to imperfections in the final measurement, the CFI is reduced by a factor $(1-2p)^4$. The second is that the effective decoherence has changed from $(2(1-p_\text{L})^{M_1}-1)^2$ to $(2(1-\tilde{p}_{\text{L},\text{p},N})(1-\tilde{p}_{\text{L},N})^{M_1}-1)^2$, representing the overhead of fault-tolerant sensing. We next compare with the CFI in the entanglement free setting, Eq.~\eqref{eq:QFIsinglequbitXnoise} in appendix~\ref{apenA4} ($\mathcal{J}\approx(2M_1p+(1-2p)^{M_1}-1)/2p^2$), which displays Heisenberg scaling up until $M_1\approx1/p$. We compare these protocols in terms of the CFI per channel use, evaluated at the optimal $M_1$. For the ancilla-free protocol, $\max_{M_1}\mathcal{J}_\text{S}/M_1\to1/p$. For the fault-tolerant protocol at $M_1=a/\tilde p_{\text{L},N}$, $\mathcal{J}/M_1=a(2\mathrm{e}^{-a}-1)^2(1-2p)^6/\tilde p_{\text{L},N}$, maximized at $a=0.1959$ where $a(2\mathrm{e}^{-a}-1)^2=0.0813$. The fault-tolerant protocol therefore wins iff $\tilde p_{\text{L},N}<0.0813\,p\,(1-2p)^6$. For $N=3,5,7,9$, the maximum values of $p$ for which there is an advantage of fault tolerant sensing are approximately $0.0002, 0.002, 0.005$, and $0.007$ respectively, as shown in Fig.~\ref{fig:FTerrorplot2}.


We next compare to a similar recent fault-tolerant sensing protocol from Ref.~\cite{sahu2026achieving}. In this protocol, only parallel sensing was considered, not sequential, and so Heisenberg scaling is in qubit number as opposed to time. Thus, given a restriction on qubit number $N$ (as is most relevant to current-day sensors), Ref.~\cite{sahu2026achieving} presents a protocol capable of achieving a Fisher information that scales as $N^2$. Hence, when allowed $M=N\nu_N$ total channel uses, the variance for parallel sensing scales as $1/(N^2\nu_N)$ for $M>N$. In contrast, given $M=M_1\nu$ total channel uses, the variance of our protocol scales as $1/(\nu M_1^2)\propto p^{2t'}/\nu=p^{(N+1)}/\nu$. Another way of saying this is that, in Ref.~\cite{sahu2026achieving}, the number of channel uses that are coherently used together is restricted by the number of qubits. In our protocol, the number of coherent channel uses is a function of both the qubit number and the noise in the channel. The limiting case of zero noise is a good example to demonstrate the differences---with sequential sensing, Heisenberg scaling can be achieved with an indefinite number of channel uses, whereas with parallel sensing Heisenberg scaling is limited to the number of qubits. Thus, although our proposal bears many similarities to Ref.~\cite{sahu2026achieving}, fault-tolerant sensing in a sequential protocol can be significantly more powerful than fault-tolerant sensing in a parallel protocol. This is demonstrated in Fig.~\ref{fig:FTerrorplot}, using logical error rates predicted by STIM~\cite{gidney2021stim}.

\subsection{Threshold in fault-tolerant quantum sensing}
The above results raise the possibility of defining a threshold in fault-tolerant quantum sensing as the value of $p$ at which $\sigma_{N,\phi}\geq \sigma_{N+2,\phi}$ for all odd $N$, where $\sigma_{N,\phi}$ is the standard deviation in estimating $\phi$ using a $N$-qubit repetition code. In Fig.~\ref{fig:FTerrorplot2}, we plot the attainable precision for different code distances. We observe a threshold at approximately $p=0.02$, below which point the attainable error decreases with increasing code distance. Note that not far above threshold the variance appears to diverge. However, this is an artifact of the fact that a worst case lower bound is used throughout this work, rather than a physical effect (see appendix~\ref{apenworstcase} for more detail).

\section{Discussion}
\label{sec:conclusion}
In this work, we have examined how techniques from fault-tolerant quantum computing can be applied to quantum sensing. Our main result, presented in Sec.~\ref{sec:FTlimitednoise}, is that, under a restricted noise model (satisfying the HNLS condition~\cite{zhou2018achieving,demkowicz2017adaptive}), Heisenberg limited quantum sensing can be achieved even when allowing for faults at every stage of the protocol. This suggests that biased noise qubits may be natural fault-tolerant quantum sensors~\cite{reglade2024quantum}. However, a great deal of further work is needed along this direction. In appendix~\ref{apen:qudit}, we extend this result to qudit systems. Whether fault-tolerant Heisenberg scaling can be achieved more broadly---for example in bosonic or fermionic systems under  similarly restricted noise models~\cite{mann2025quantum,ghosh2025optimal,bravyi2018correcting,marton2023coherent,iverson2020coherence,huang2019performance}, remains unclear. These results provide some direction as to the application of fault-tolerant techniques to quantum sensing, however a more general connection is still lacking. Although there may be no reason to expect such a connection to exist, we are motivated by the fact that fault-tolerant techniques have been demonstrated to be beneficial for quantum communication~\cite{ye2015fault,christandl2022fault,belzig2024fault,shi2025stabilizer,shi2025measurement,christandl2024fault}. Indeed, there is a well-established fundamental connection between entanglement purification and error correction~\cite{bennett1996mixed}. Whether such a connection for sensing exists beyond that established in Refs.~\cite{zhou2018achieving,demkowicz2017adaptive,kubica2021using} is as of yet unclear. 



On a positive note, there remains a huge number of open questions to be answered. In this work, we have made a number of potentially avoidable assumptions. In Sec.~\ref{sec:FTlimitednoise}, we assumed that $\phi$ is approximately known in advance and is sufficiently small (see appendix~\ref{apen:errordet}), that our estimators must be unbiased, and that our model is perfectly known. Relaxing any of these assumptions would create open problems potentially relevant for practical settings. Additionally, the results in Sec.~\ref{sec:FTlimitednoise} could be extended to multiparameter estimation~\cite{omanakuttan2024quantum,gorecki2020optimal}, continuous variable systems~\cite{zhuang2020distributed}, or non-unitary encodings~\cite{wagner2021optimal,wagner2022pauli}. Merging our results with those of Ref.~\cite{sahu2026achieving} to achieve Heisenberg scaling in both time and qubit number is a natural extension.


Several other directions appear promising. In Ref.~\cite{zaiser2016enhancing}, enhanced quantum sensing performance was achieved via a quantum memory. Extending this to the case of a logical quantum memory may provide a further increase in performance. Re-examining the settings considered in this work, the case with
adversarial noise presents another potential regime where fault-tolerant sensors may see an enhancement over conventional sensors. Additionally, it may be informative to study the estimation of logical rotations, even if this does not have direct applicability to real-world sensing problems. There are several different ways of implementing logical rotations~\cite{huang2025robust} which may demonstrate different performance.

Finally, the most promising avenue (as we see it) for future studies in fault-tolerant quantum sensing arises from the biggest issue faced. At the heart of the incompatibility between quantum sensing and fault-tolerance lies the Eastin--Knill theorem~\cite{EastinKnill2009}. This states that no finite-dimensional QEC code can have a universal transversal logical gate set. This constraint flies in the face of quantum sensing, where the typical problem set-up is to estimate a continuous parameter encoded through some transversal operation. Recently, Ref.~\cite{ott2026rare} has presented one regime in which logical qubits can be useful in sensing despite this limitation.
However, whether this apparent incompatibility can be overcome to achieve a scalable advantage remains unclear. One potentially promising avenue for this is the use of approximate quantum error correcting codes, although there are limitations in this setting also~\cite{faist2020continuous,kubica2021using,liu2023approximate}. On the other hand, known code properties might be useful for determining fundamental limits on fault-tolerant sensing~\cite{bravyi2012magic,rengaswamy2020optimality}.






\section*{Acknowledgements}
We acknowledge useful discussions with Animesh Datta, Jacob Lin, and Sisi Zhou. 
L.O.C.~was supported in part by 
the National Science Foundation under Award No. 2533041. 
L.O.C., E.A.~and A.V.G.~were supported in part by ONR MURI, AFOSR MURI, NSF QLCI (award No.~OMA-2120757),  NSF STAQ program, DoE ASCR Quantum Testbed Pathfinder program (award No.~DE-SC0024220),  ARL (W911NF-24-2-0107), DARPA SAVaNT ADVENT, and NQVL:QSTD:Pilot:FTL. L.O.C., E.A.~and A.V.G.~also acknowledge support from the U.S.~Department of Energy, Office of Science, National Quantum Information Science Research Centers, Quantum Systems Accelerator (award No.~DE-SCL0000121) and from the U.S.~Department of Energy, Office of Science, Accelerated Research in Quantum Computing, Fundamental Algorithmic Research toward Quantum Utility (FAR-Qu).
Y.-X.W.~acknowledges support from a QuICS Hartree Postdoctoral Fellowship.

\appendix

\section{Ancilla-free bounds}
\label{apenA4}
In this appendix, we present simple bounds on the attainable precision for estimating a rotation about the $Z$ axis under different noise models when no ancilla qubits are used. Equivalently, these are single-qubit bounds. In Secs.~\ref{subsec:signoisesame} and \ref{subsec:signoisediff}, respectively, we consider the signal and noise along the same and different directions.

\subsection{Signal and noise along same direction}
\label{subsec:signoisesame}
First consider the simple case of estimating a rotation about the $Z$ axis in the presence of $Z$ noise. Starting from the state $\ket{+}$, solving Eq.~\eqref{eq:lindblad_metrology}, we find that the quantum state after a time $t$ is given by
\begin{equation}
    \rho(t)=\frac{1}{2}\begin{pmatrix}
        1&\mathrm{e}^{-2t\gamma-it\phi}\\
        \mathrm{e}^{-2t\gamma+it\phi}&1
    \end{pmatrix}\;.
\end{equation}
Computing the corresponding QFI one finds \mbox{$\mathcal{J}_\text{S}=\mathrm{e}^{-4\gamma t}t^2$}. Given a total sensing time $T$, one can repeat the experiment $T/t$ times, hence we are interested in maximizing $\mathcal{J}_\text{S}T/t$. This is optimized when $t=1/(4\gamma)$ as expected. Therefore, given a total sensing time $T$, Heisenberg scaling can be achieved for $T=O(1/\gamma)$~\cite{Huelga1997}.

The discrete-time analog of this channel is a $Z$ operation implemented with probability $p$, followed by a rotation around the $Z$ axis, leaving the $\ket{+}$ state in the following state:
\begin{equation}
    \rho=\frac{1}{2}\begin{pmatrix}
        1&\mathrm{e}^{-i\phi}(1-2p)\\
        \mathrm{e}^{i\phi}(1-2p)&1
    \end{pmatrix}\;.
\end{equation}
After $M_1$ uses of the channel, the state evolves to 
\begin{equation}
\label{eq:rhoapenA4}
    \rho=\frac{1}{2}\begin{pmatrix}
        1&\mathrm{e}^{-iM_1\phi}(1-2p)^{M_1}\\
        \mathrm{e}^{i{M_1}\phi}(1-2p)^{M_1}&1
    \end{pmatrix}\;.
\end{equation}
The QFI can be computed as $\mathcal{J}_\text{S}={M_1}^2(1-2p)^{2M_1}$. Given $M=\nu M_1$ total channel uses, we wish to maximize $\mathcal{J}_\text{S}M/M_1$. This expression is maximized when $M_1=-1/[2\log(1-2p)]\approx1/(4p)$. This gives Heisenberg scaling in $M$ up until $O(1/p)$, the analog of Heisenberg scaling up until $T=O(1/\gamma)$ above.

We now demonstrate a measurement saturating the expression above. Measuring $\rho$ in Eq.~\eqref{eq:rhoapenA4} in the $Y$ basis, we find probabilities of
\begin{equation}
\label{eq:probabilitysinglequbit}
    p_\pm=\frac{1}{2}(1\pm(1-2p)^{M_1}\sin(M_1\phi))\;,
\end{equation}
where $p_\pm=\text{Tr}[\rho\ket{\pm_y}\bra{\pm_y}]$.
Computing the CFI via Eq.~\eqref{eq:CFIdefin}, one finds
\begin{equation}
    \mathcal{J}=\frac{M_1^2 }{(1-2 p)^{-2 M_1}\sec(M_1\phi)^2-\tan(M_1\phi)^2 }\;.
\end{equation}
Operating near $\phi=0$, one finds $\mathcal{J}=M_1^2(1-2p)^{2M_1}$, in agreement with the QFI.

Alternatively, one can apply $R_z(-M_1\phi-\pi/2)$ to the probe state and then measure in the $X$ basis to get the same results. The $\phi$ dependence of these optimal measurements is well established in quantum metrology literature, and can be implemented via an adaptive protocol~\cite{barndorff2000fisher,hayashi2005statistical}.


\subsection{Signal and noise along different directions}
\label{subsec:signoisediff}
The single-qubit version of the channel considered in Sec.~\ref{sec:FTlimitednoise} is estimating a rotation about the $Z$ axis in the presence of $X$ noise. Solving Eq.~\eqref{eq:lindblad_metrology} with $L_x=\sqrt{\gamma}X$, we find that the $\ket{+}$ state evolves to
\begin{widetext}
   \begin{equation}
    \rho(t)=
\frac{1}{2}\begin{pmatrix}
  1&  e^{-\gamma t} \left(\cosh \left(t \sqrt{\gamma ^2-\phi ^2}\right)+\frac{(\gamma -i \phi ) \sinh \left(t \sqrt{\gamma ^2-\phi ^2}\right)}{\sqrt{(\gamma -\phi ) (\gamma +\phi )}}\right) \\
 e^{-\gamma t} \left(\cosh \left(t \sqrt{\gamma ^2-\phi ^2}\right)+\frac{(\gamma +i \phi ) \sinh \left(t \sqrt{\gamma ^2-\phi ^2}\right)}{\sqrt{(\gamma -\phi ) (\gamma +\phi )}}\right) & 1 
\end{pmatrix}\;.
\end{equation} 
\end{widetext}
Taking the limit $\phi\to0$, we can compute the QFI as
\begin{equation}
\label{eq:apenA7}
\mathcal{J}_\text{S}=\frac{2t\gamma+e^{-2 \gamma  t} -1}{2\gamma ^2}\;.
\end{equation}
We note that this expression is computed using Refs.~\cite{dittmann1999explicit,liu2020quantum}, and this example demonstrates a discontinuity in the QFI, when the rank of the state changes (i.e. exactly at $\phi=0$)~\cite{vsafranek2017discontinuities}. Observe that, for $t\ll1/\gamma$, $\mathcal{J}_\text{S}$ scales like $t^2$ as expected, and for $t\gg1/\gamma$, $\mathcal{J}_\text{S}$ goes as $t/\gamma$. This shows a crossover from quadratic to linear growth at the characteristic timescale $t\approx1/\gamma$.

In this example, given a total sensing time $T$, the optimal value of $t$ is $T$. For $T=1/\gamma$, this gives Heisenberg scaling in $T$. For $T\gg 1/\gamma$, this gives $\mathcal{J}_\text{S}\approx T/\gamma$, i.e.~SQL in $T$. This is in agreement with known results~\cite{zhou2024limits}. 


For constructing an effective model where we interact with a channel $M_1$ times with a bit-flip probability $p$ per channel use, we can take $t=M_1\delta t$, and $\mathrm{e}^{-2\gamma t}=(\mathrm{e}^{-2\gamma \delta t})^{M_1}=(1-2p)^{M_1}$. As $\delta t$ is unimportant (can be absorbed into $\phi$ for example) and is fixed, we set $\delta t=1$ to normalize the phase acquired per channel use. Then substituting $\gamma=-\log(1-2p)/2$, and $t=M_1$, we arrive at the following expression for the QFI to leading order in $p$
\begin{equation}
\label{eq:QFIsinglequbitXnoise}
\begin{split}
    \mathcal{J}_\text{S}&=\frac{2 \left((1-2 p)^{M_1}-M_1 \log (1-2 p)-1\right)}{\log(1-2 p)^2}\\&\approx\frac{2M_1p+(1-2p)^{M_1}-1}{2p^2}
    \end{split}
\end{equation}
We focus on $0 \le p \le 1/2$ as this is the physically relevant regime. 


As above, this shows Heisenberg scaling in $M_1$ ($ \mathcal{J}_\text{S}\approx M_1^2$) for $M_1\ll1/p$. For $M_1\gg 1/p$, we find $ \mathcal{J}_\text{S}\approx M_1/p$, i.e.~SQL in $M_1$. There is therefore a similar transition from Heisenberg scaling to the SQL scaling when $M_1=O(1/p)$.

Importantly, this shows that to within constant terms bit-flip and phase-flip noise models give the same sensitivity at the single-qubit level. 


\subsubsection{Optimality of $\ket{+}$ state}
Above, we have used the $\ket{+}$ state as our probe state. We now examine the conditions under which this is the optimal single-qubit probe state to use. Starting from $\rho=I_2/2+(s_x X+s_y Y+s_z Z)/2$, the temporal derivative is given by
\begin{equation}
    \dot{\rho}=\left(
\begin{array}{cc}
 -\gamma s_z & -\frac{1}{2} \phi(is_x  + s_y  )+i\gamma s_y \\
 \frac{1}{2} \phi  (i s_x-s_y)-i \gamma  s_y & \gamma  s_z \\
\end{array}
\right)\;.
\end{equation}
Since $s_z$ obeys a closed equation, $\dot{s}_z=-2\gamma s_z$, it evolves independently of $s_x$ and $s_y$ and cannot be converted into useful $\phi$ sensitivity. Thus the optimal single-qubit probe may be taken to satisfy $s_z=0$. 


We therefore arrive at two equations from the real and imaginary parts of $\rho$:
\begin{equation}
    \begin{split}
        \dot{s}_x&=-\phi s_y\;,\\
        \dot{s}_y&=-2s_y\gamma+\phi s_x\;.
    \end{split}
\end{equation}
Differentiating $ \dot{s}_x$ again and eliminating $s_y(t)$, we obtain
\begin{equation}
    \ddot{s}_x+2\dot{s}_x\gamma+\phi^2 s_x =0\;.
\end{equation}
This has a solution
\begin{equation}
    s_x(t)=c_1 e^{t \left(-\sqrt{\gamma ^2-\phi ^2}-\gamma \right)}+c_2 e^{t \left(\sqrt{\gamma ^2-\phi ^2}-\gamma \right)} \;.
\end{equation}
Using $s_y=-\dot{s}_x/\phi$, we arrive at
\begin{widetext}
    \begin{equation}
    s_y(t)=-\frac{ \left(c_2 \left(\sqrt{\gamma ^2-\phi ^2}-\gamma \right) e^{ -t (-\sqrt{\gamma ^2-\phi ^2}+\gamma)}-c_1 \left(\sqrt{\gamma ^2-\phi ^2}+\gamma \right)e^{-t \left(\sqrt{\gamma ^2-\phi ^2}+\gamma \right)}\right)}{\phi }\;.
\end{equation}
\end{widetext}
For these equations to make sense $s_x(0)$ and $s_y(0)$ should be independent of $\phi$. Observe that 
\begin{equation}
\begin{split}
s_x(0)&=c_1+c_2,\\
        s_y(0)&=-\frac{c_2 \left(\sqrt{\gamma ^2-\phi ^2}-\gamma \right)-c_1 \left(\sqrt{\gamma ^2-\phi ^2}+\gamma \right)}{\phi }\;.
\end{split}
\end{equation}
Solving such that $s_x(0)=s_{x,0}$ and $s_y(0)=s_{y,0}$ are independent of $\phi$, we find
\begin{equation}
\begin{split}
    c_1=&\frac{s_{x,0}\left(\sqrt{\gamma ^2-\phi ^2}-\gamma \right)+s_{y,0} \phi }{2 \sqrt{\gamma^2 -\phi^2}},\\
        c_2=&\frac{s_{x,0} \left(\sqrt{\gamma ^2-\phi ^2}+\gamma \right)-s_{y,0} \phi }{2 \sqrt{\gamma^2-\phi^2}}\;.
\end{split}
\end{equation}
From the convexity of the QFI we can restrict the optimal state to be pure, taking $s_{x,0}=\cos(\theta)$ and $s_{y,0}=\sin(\theta)$. Substituting these expressions in, we find
\begin{widetext}
    \begin{equation}
    s_x(t)=e^{-\gamma t } \left(\cos (\theta ) \cosh \left(t \sqrt{\gamma ^2-\phi ^2}\right)+\frac{\sinh \left(t \sqrt{\gamma ^2-\phi ^2}\right) (\gamma  \cos (\theta )-\phi  \sin (\theta ))}{\sqrt{(\gamma -\phi ) (\gamma +\phi )}}\right)\;,
\end{equation}
and
\begin{equation}
    s_y(t)=e^{-\gamma t} \left(\sin (\theta ) \cosh \left(t \sqrt{\gamma ^2-\phi ^2}\right)+\frac{\sinh \left(t \sqrt{\gamma ^2-\phi ^2}\right) (-\gamma  \sin (\theta )+\phi  \cos (\theta ))}{\sqrt{(\gamma -\phi ) (\gamma +\phi )}}\right)\;.
\end{equation}
\end{widetext}
From this we are able to compute the QFI using Refs.~\cite{dittmann1999explicit,liu2020quantum}, and we observe that the optimal $\theta$ is indeed $\phi$-dependent, meaning the $\ket{+}$ state is not generically optimal. Additionally, note that taking the limits $\theta\to0$ and $\phi\to0$ do not commute for this example. Taking the limit $\theta\to0$ first, followed by $\phi\to0$ recovers Eq.~\eqref{eq:apenA7}. In contrast, taking the limit $\phi\to0$, we find that the QFI is optimized as $\theta\to0$ as expected, and is given by
\begin{equation}
   \mathcal{J}_\text{S}= \frac{e^{-2 \gamma  t} \sinh ^2(\gamma  t) \left( \tanh (\gamma  t)+1\right)}{\gamma ^2}\;.
   \label{eq:A18}
\end{equation}
We stress that the disagreement with Eq.~\eqref{eq:apenA7} is well understood~\cite{vsafranek2017discontinuities}. Note that, for every nonzero $\phi$ the achievable QFI is Eq.~\eqref{eq:apenA7}, which is larger than Eq.~\eqref{eq:A18}. The above implies that the $\ket{+}$ probe is asymptotically optimal in the local regime $\phi\to0$, even though it is not generically optimal for fixed non-zero $\phi$. In the regime of interest for this paper ($\phi\ll1$), or in an adaptive protocol, where $\phi$ is learned bit by bit, and control rotations cancel the known rotation, the effective rotation will become smaller, and $\ket{+}$ becomes asymptotically optimal.

\section{Fault-tolerant sensing strategy}
\label{apen:faultolgeneral}
In this appendix, we present the details of our fault-tolerant sensing strategy. In Sec.~\ref{apen:unbiasedest}, we discuss how to construct an unbiased estimator based on the observed syndrome measurement results. We then describe the impact of logical errors in the ideal error correction setting in Sec.~\ref{apen:problemreduction}. In Secs.~\ref{apen:stateprep}, \ref{apen:errordet}, and \ref{apen:FTmeas}, we discuss the fault-tolerant state preparation, syndrome extraction and correction, and measurement required for a 3-qubit repetition code, respectively. In Sec.~\ref{apen:nqubitrepetition}, this is generalized to the $N$-qubit repetition code. Finally, in Sec.~\ref{apen:zerrors}, we discuss the impact of phase-flip errors.

\subsection{Unbiased estimator based on observed syndromes}
\label{apen:unbiasedest}
Here we describe how to construct an unbiased estimator for the setting in Sec.~\ref{subsec:idealsetting}. To see the reason this is necessary, consider a 3-qubit repetition code. After 2 rounds, if no errors are detected, one might be tempted to assume we have the ideal GHZ state $\mathrm{e}^{-i\phi}\ket{0}^{\otimes3}+\mathrm{e}^{i\phi}\ket{1}^{\otimes3}$. By uncomputing the GHZ state, and measuring in the $X$-basis to obtain probabilities $(1\pm\cos(2\phi))/2$, one could estimate $\phi$. However, observe that in the setting described, with probability $p^3((1-p)^3+p^3)$, the total phase picked up after 2 rounds is 0, not $2\phi$. The factor $p^3$ comes from all 3 qubits experiencing a bit-flip in between the first and second channels. The factor $((1-p)^3+p^3)$ is the probability of either 0 qubits or 3 qubits experiencing a bit flip during the first round. Not accounting for this would result in an estimator with magnitude too small by a constant factor. An unbiased estimator must account for this possibility.

Returning to the $N$-qubit repetition code, let $\Phi$ denote the accumulated phase over all observed syndromes $\mathbf{s}$. Then
\begin{equation}
    p_i(\mathbf{s})
=
\Pr\!\left[
\Phi=(M_1-2i)\phi\mid\mathbf{s}
\right]
\end{equation}
denotes the conditional probability of each accumulated phase given the
observed syndrome history $\mathbf{s}$. The probabilities $p_i(s)$ are obtained by summing the probabilities of all physical error histories compatible with the observed syndrome record $s$ that produce exactly $i$ negative phase contributions, and normalizing by the total probability of observing $s$. We choose the final logical
measurement basis so that, for a definite accumulated phase $\Phi$, the
outcome $x\in\{-1,+1\}$ has mean $\sin\Phi$. Equivalently, this is a
logical-$Y$ measurement, or a logical-$X$ parity measurement preceded by
a known $\pi/2$ phase shift. The conditional measurement probabilities
are then
\begin{equation}
\Pr(x\mid\mathbf{s},\phi)
=
\frac{1}{2}\left[
1+xg_{\mathbf{s}}(\phi)
\right],
\end{equation}
where
\begin{equation}
g_{\mathbf{s}}(\phi)
=
\sum_{i=0}^{M_1}
p_i(\mathbf{s})
\sin[(M_1-2i)\phi].
\end{equation}
For small $\phi$,
\begin{equation}
g_{\mathbf{s}}(\phi)
=
c(\mathbf{s})\phi+O(\phi^3),
\qquad
c(\mathbf{s})
=
\sum_{i=0}^{M_1}
p_i(\mathbf{s})(M_1-2i).
\end{equation}

After $\nu$ independent repetitions, with records
$(\mathbf{s}_r,x_r)$, define
\begin{equation}
\hat{\phi}
=
\frac{\sum_{r=1}^{\nu}c(\mathbf{s}_r)x_r}
{\sum_{r=1}^{\nu}c(\mathbf{s}_r)^2}.
\end{equation}
Conditioned on the observed syndrome histories (using $\mathbb{E}[x_r\mid\mathbf{s}_r]=c(s_r)\phi+O(\phi^3)$),
\begin{equation}
\mathbb{E}[\hat{\phi}\mid\mathbf{s}_1,\ldots,\mathbf{s}_\nu]
=
\phi+O(\phi^3),
\end{equation}
so $\hat{\phi}$ is locally unbiased at $\phi=0$. The variance of this measurement strategy is given by
\begin{equation}
    \mathbb{E}\left[\left(\hat{\phi}-\mathbb{E}[\hat{\phi}]\right)^2\right] \approx1/(\nu\,\mathbb{E}[c(\mathbf{s})^2])\;.
\end{equation}
This is efficient, in that it saturates the CFI, and is bounded in the next section.

\subsection{Impact of logical errors on attainable precision}
\label{apen:problemreduction}
Recall Eq.~\eqref{eq:GHZideal} that describes the state obtained after $M_1$ rounds of interacting with the channel:
\begin{widetext}
\begin{equation}
\label{eq:GHZideal2}
    \rho_{\text{GHZ},M_1}(\phi)=(1-p_\text{L})^{M_1}\ket{\psi_\text{GHZ}(\phi_{M_1})}\bra{\psi_\text{GHZ}(\phi_{M_1})}+(1-(1-p_\text{L})^{M_1})\tilde{\rho}\;.
\end{equation}
\end{widetext}
We now present two ways of bounding the decoherence introduced by logical errors. As a simplified model, one can take the continuous time limit, i.e.~follow Eq.~\eqref{eq:lindblad_metrology}, where jump operators are modeled as logical $X$ operations. A GHZ state with jump operators as logical $X$ operations follows the same equations of motion as the single-qubit case discussed in Sec.~\ref{subsec:signoisediff} above, with $\gamma$ replaced by the logical decoherence rate. Equivalently, the discrete-time-step analog from above shows that the QFI is given by Eq.~\eqref{eq:QFIsinglequbitXnoise} with $p$ replaced by $p_\text{L}$. While this simplified model provides some intuition for the expected result, it does not capture the full dynamics of the noise model of interest.

Alternatively, we can examine the measurement statistics from an explicit measurement protocol. One option is to measure in the logical $X$ basis. To do this, we apply a Hadamard to each qubit and measure in the computational basis. Doing so reveals that the positive parity outcomes are observed with probability $(1+(2(1-p_\text{L})^{M_1}-1)\cos(\phi_{M_1}))/2^N$ for a $N$-qubit GHZ state. Similarly, negative parity outcomes are observed with probability $(1-(2(1-p_\text{L})^{M_1}-1)\cos(\phi_{M_1}))/2^N$. Note that here we have made the conservative assumption that, when a logical error occurs, the output state has phase of $\phi_{M_1}+\pi$. \footnote{This is conservative for this measurement because it minimizes the CFI by making the parity signal from the error branch oppose that of the ideal branch.} (From the previous subsection we know that the true distribution of $\phi_{M_1}$ will be syndrome dependent.) It is important to stress that this approximation is valid when $(1-p_\text{L})^{M_1}\approx1$, i.e. when $M_1\ll1/p_\text{L}$. This is because in this regime the contribution from $\tilde{\rho}$ is negligible, hence any assumption about the impact of logical errors will not change the estimation error in any meaningful way. Combining all the positive parity outcomes into a single measurement outcome, and similarly for the negative parity, leaves us with two probabilities:
\begin{equation}
    p_\pm=\frac{1}{2}(1\pm(2(1-p_\text{L})^{M_1}-1)\cos(\phi_{M_1}))\;.
\end{equation}
The corresponding CFI is
\begin{equation}
    \mathcal{J}=-\frac{(2(1-p_\text{L})^{M_1}-1)^2 M_1^2 (1-2 p)^2 \sin ^2(M_1 (1-2 p) \phi )}{(2(1-p_\text{L})^{M_1}-1)^2 \cos ^2(M_1 (1-2 p) \phi )-1}\;.
\end{equation}
Evaluating at $\phi=\pi/(2M_1(1-2p))$ (which can be done via an adaptive protocol~\cite{barndorff2000fisher,hayashi2005statistical}) gives a CFI of
\begin{equation}
\label{eq:QFIlogical}
    \mathcal{J}=(2(1-p_\text{L})^{M_1}-1)^2 M_1^2 (1-2 p)^2\;.
\end{equation}
Note that this is equivalent to the measurement considered in appendix~\ref{apen:unbiasedest} and appendix~\ref{apenworstcase}, up to a rotation. This achieves Heisenberg scaling in $M_1$, provided $M_1\ll1/p_\text{L}$. Note that this restriction does not remove the possibility of achieving Heisenberg scaling for larger $M_1$ by increasing the code size, as by writing $p_\text{L}\approx c_{t'}p^{t'}$, below threshold we can always find a $t'$ such that $1/p\ll M_1\ll 1/(c_{t'}p^{t'})$.

For completeness, we now demonstrate that the above assumptions and the restriction $M_1\ll1/p_\text{L}$ can be avoided. To remove the restriction $M_1\ll1/p_\text{L}$, we must first understand its origin. Observe that \mbox{$(2(1-p_\text{L})^{M_1}-1)^2\to1$} as $M_1\to\infty$. At first glance, this appears odd, as it suggests that as $M_1\to\infty$ there is no decoherence effect. However, this is only an artifact of the fact that we modeled $\tilde{\rho}$ as carrying the phase $\phi_{M_1}+\pi$. This is a fine (and conservative) assumption provided $M_1\ll1/p_\text{L}$ is satisfied. More accurately however, $\tilde{\rho}$ will be a mixture of states with a distribution of phase values ranging from $-M_1\phi$ to $(M_1-1)\phi$. As such, in appendix~\ref{apenworstcase}, we present a worst case bound for states of the form shown in Eq.~\eqref{eq:GHZideal2}. This bound demonstrates that even under the worst case distribution of acquired phases, the bound in Eq.~\eqref{eq:QFIlogical} is valid for any $\frac{1}{2}\leq(1-p_\text{L})^{M_1}\leq 1$ (Note that the bound in appendix~\ref{apenworstcase} omits an additional factor $(1-2p)^2$ arising from the reduction in signal). Eq.~\eqref{eq:QFIlogical} is maximized for $M_1\approx-0.315/\log(1-p_\text{L})$. We stress that while it is possible to derive an exact bound on the QFI for the state in Eq.~\eqref{eq:GHZideal2}, the difficulty is that the state $\tilde{\rho}$ is dependent on the observed syndromes across all $M_1$ rounds of the experiment. Thus, given a set of observed syndromes it is feasible to compute the corresponding $\tilde{\rho}$. However, computing $\tilde{\rho}$ averaged over all possible observed syndromes seems challenging. As such, we use the worst case bounds above.

An alternative measurement is to uncompute the GHZ circuit mapping the state in Eq.~\eqref{eq:GHZideal} to the single-qubit equivalent. From there the standard single-qubit measurement described in appendix~\ref{apenA4} achieves the same CFI as above.



\subsection{Fault-tolerant state preparation}
\label{apen:stateprep}
We now move on to describe our fault-tolerant sensing protocol, that extends the regime over which Heisenberg scaling can be achieved, even when allowing for errors after every operation as described in Sec.~\ref{subsec:circuitlevel}. For simplicity, we begin by describing a three-qubit repetition code, with the $N$-qubit code described in appendix~\ref{apen:nqubitrepetition}. The first stage of this is state preparation, where we use an ancilla qubit to check whether our state was prepared correctly. Observe that an $X$ error before the first Hadamard gate can propagate through to create the $\ket{000}-\ket{111}$ state. The simplest way to prevent such errors is to measure the state of the first qubit after the first Hadamard. After the Hadamard the first qubit is either in the state $\ket{+}$ or $\ket{-}$. By preparing an ancilla in the $\ket{+}$ state, acting a CNOT from the ancilla to the qubit (controlled on the ancilla), performing a Hadamard on the ancilla and measuring in the computational basis, we can determine the state of the first qubit. After performing the corresponding correction operation, this offers a simple method of suppressing any errors in the phase of the initial GHZ state to $O(p^2)$.

Hence, going forward we focus on $X$ errors. After creating the GHZ state with fidelity $1-O(p)$, we implement a CNOT from qubit 1 to the ancilla (controlled on qubit 1). We next implement a CNOT from qubit 2 to the ancilla. If the ideal GHZ state was created, the ancilla is in the $\ket{0}$ state. However, if either $\ket{100}+\ket{011}$ or $\ket{010}+\ket{101}$ was created, the final ancilla is in the $\ket{1}$ state. We then measure to determine $Z_1Z_2$. By repeating and measuring $Z_2Z_3$, the location of the error can be determined, and either discarded or corrected.


However, observe that in both of the above cases, if the ancilla qubit also experiences $X$ errors, with probability $p$ (either in state preparation or measurement), it can appear as though there is an $X$ error on an ideal GHZ.
We therefore need to repeat the syndrome measurement three times, and we only keep the state if all measurements agree and show that no errors are present (Here we are assuming the probability of an error in the syndrome measurement is $p$. For a more general discussion on how many syndrome measurements are needed see Sec.~\ref{apensyndromerepetitions}.). Alternatively, deterministic state preparation can be achieved as follows. Based on the three observed syndrome measurements, a decision can be made on what error was most likely to have occurred. By correcting the appropriate error, we can achieve state preparation with logical error probability $O(p^2)$.

\begin{figure}[t!]
    \centering
\includegraphics[width=0.95\columnwidth]{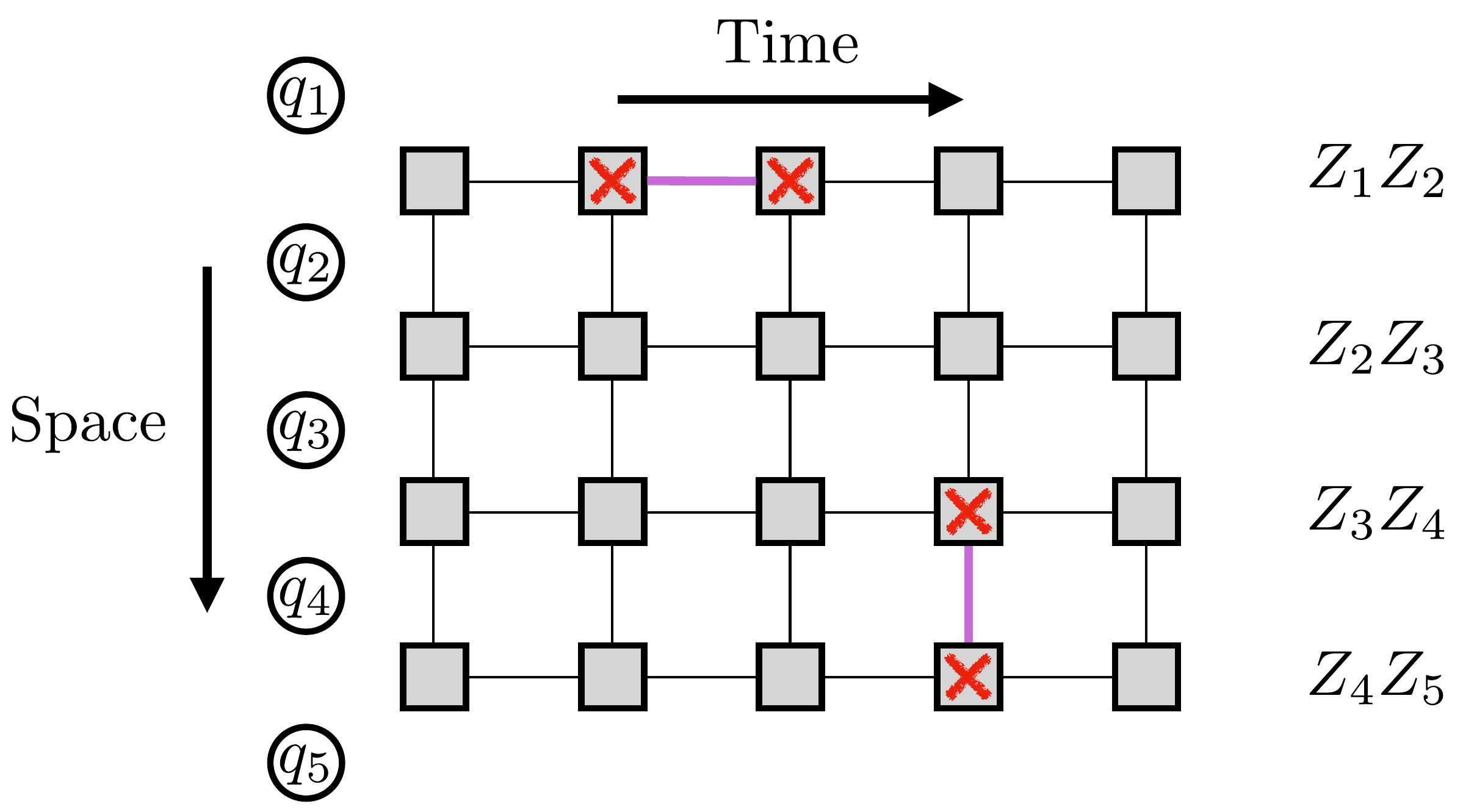}
    \caption{\textbf{Sample detection events and the minimum-weight-perfect-matching decoder.} Grey boxes correspond to detection events, i.e. the difference between consecutive syndrome measurement results modulo 2. Red crosses show detection events that differ from the noiseless case. Qubits are labeled as $q_i$, and the measured syndrome for each row $Z_iZ_{i+1}$ is shown. We show two types of errors and the corresponding decoding. The horizontal purple line arises from a detector error, whereas the vertical purple line corresponds to an error on qubit 4. }
    \label{fig:MWPM}
\end{figure}

\subsection{Fault-tolerant syndrome extraction and correction}
\label{apen:errordet}
After the fault-tolerant state preparation above, the ideal GHZ state has been prepared to within a correctable error with probability $1-O(p^2)$. The next stage is to interact with the sensing channel, which introduces $X$ errors on every qubit with probability $p$ and also implements a rotation about the $Z$ axis on the first qubit. We look for $X$ errors in the same manner as above, detecting them with probability $1-O(p^2)$. If no errors happened in the channel, then we correctly identify this with probability $1-O(p^2)$. If one error happens with probability $O(p)$, we also correctly identify this with probability $1-O(p^2)$. Specifically, we use minimum weight perfect matching as our decoder~\cite{higgott2025sparse,demarti2024decoding}, although many other decoding techniques are known~\cite{demarti2024decoding,iyer2015hardness,bravyi2014efficient,torlai2017neural,higgott2023improved}.


To gain some understanding of how this works for a simple example, we return to Fig.~\ref{fig:FTsyndromeextract} showing the measured syndromes. For a five-qubit repetition code we measure the syndromes $s_{12}=Z_1Z_2$, $s_{23}=Z_2Z_3$, $s_{34}=Z_3Z_4$ and $s_{45}=Z_4Z_5$ (We use the five-qubit code here for clarity). In an ideal setting with no errors, all of these syndromes would return the value +1. With errors, some of the syndromes may return the value $-1$. Mapping $\pm1$ to $0,1$, we then define a detection event as the difference between consecutive syndrome measurements modulo 2, e.g.~$D_{12}(t_i)=s_{12}(t_i)-s_{12}(t_{i-1}) \mod 2$. Minimum weight perfect matching then finds the minimum weight error that is consistent with the observed detection events, shown in Fig.~\ref{fig:MWPM}.  If all errors are equally likely, then the minimum weight error is simply the shortest path between two detection events. More generally, different weights can be assigned to the different errors~\cite{dennis2002topological}. The number of times syndrome measurements are repeated is discussed in Sec.~\ref{apensyndromerepetitions}.

\begin{figure}[t]
    \centering
\includegraphics[width=0.9\columnwidth]{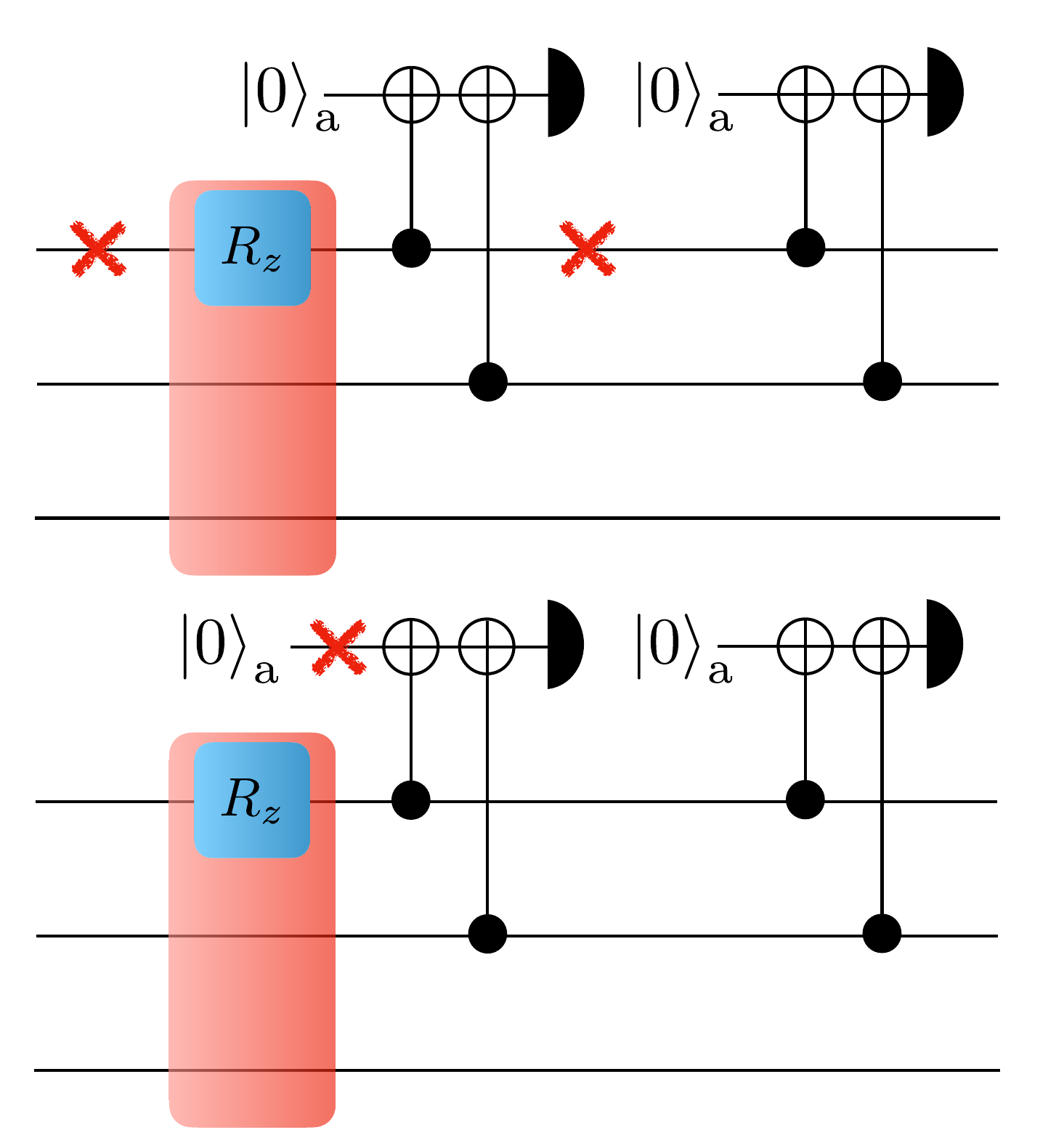}
    \caption{\textbf{Errors giving rise to equivalent syndrome measurement results.} Top and bottom figures show two different error mechanisms that appear equivalent from the viewpoint of the syndrome measurements. }
    \label{fig:equivsyndr}
\end{figure}

We now discuss a subtlety that arises from this method of syndrome extraction. As shown in Fig.~\ref{fig:equivsyndr}, there are different error mechanisms that look equivalent to the decoder. Neither of the errors shown in Fig.~\ref{fig:equivsyndr} will result in any correction operations being applied (assuming a third perfect measurement). However, crucially, the top and bottom errors in Fig.~\ref{fig:equivsyndr} correspond to acquiring a phase of $+\phi$ and $-\phi$ respectively during this sensing round. Note that this is fundamentally different to a logical error happening elsewhere in the code, which flips the sign of any previously acquired phase, and occurs with probability $\tilde{p}_{\text{L},3}$. If unaccounted for, this effect could potentially eliminate any Heisenberg scaling. However, in the next subsubsection, we verify that this effect contributes negligibly to the QFI.

\subsubsection{Mis-identified phase errors}
\label{subsubsecnonlogicalerrs}

\begin{figure}[t]
    \centering
\includegraphics[width=0.9\columnwidth]{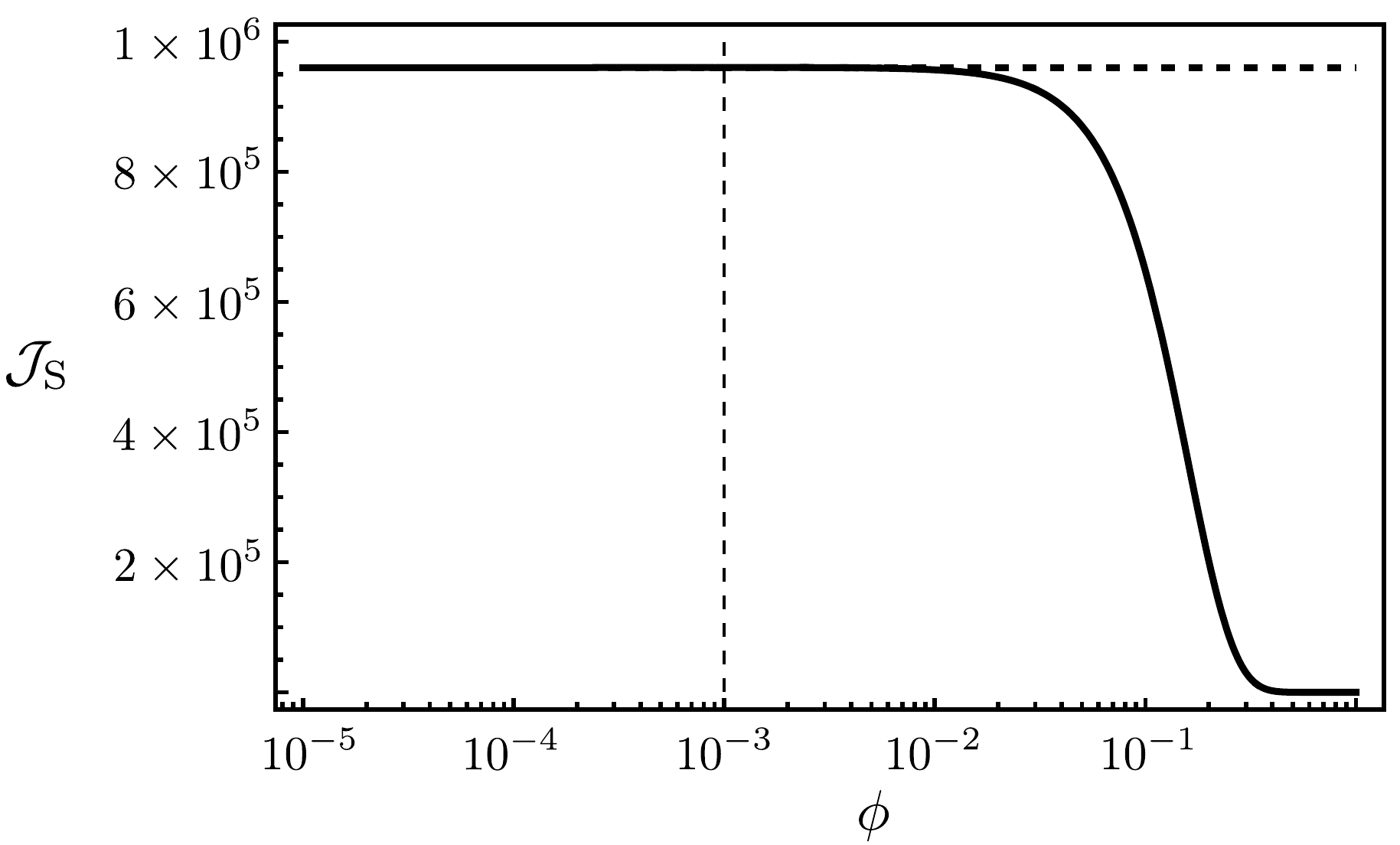}
    \caption{\textbf{Quantum Fisher information for the channel described in Eq.~\eqref{eq:channelnonlog}.} The solid black line shows the QFI for estimating $\phi$ under a channel that with probability $p^2$ implements a rotation of $-\phi$. The dashed horizontal line shows the QFI evaluated in the limit $\phi\to0$ (Eq.~\eqref{eq:QFIphi0}). The dashed vertical line corresponds to $\phi=1/M_1$, for $M_1=1000$. $p=0.1$ for this plot}
    \label{fig:equivsyndrQFI}
\end{figure}

We can describe the above effect as a quantum channel that, with probability $1-p^2$, implements the desired rotation $R_z(\phi)$, and with probability $p^2$ implements the rotation $R_z(-\phi)$:
\begin{equation}
\label{eq:channelnonlog}
\mathcal{E}(\rho)=(1-p^2)R_z(\phi)\rho R_z(\phi)^\dagger + p^2R_z(-\phi)\rho R_z(-\phi)^\dagger\;.
\end{equation}
We analyze this channel for a single-qubit state as the results directly carry over to the GHZ state of interest. Starting from the $\ket{+}$ state after $M_1$ channel uses, the state becomes
\begin{widetext}
\begin{equation}
\label{eq:apenB4rho}
    \frac{1}{2}\begin{pmatrix}
        1&(\cos(\phi)-i(1-2p^2)\sin(\phi))^{M_1}\\
        (\cos(\phi)+i(1-2p^2)\sin(\phi))^{M_1}&1
    \end{pmatrix}\;.
\end{equation}    
\end{widetext}
From this, the QFI can be computed and is found to be (evaluated as $\phi\to0$)
\begin{equation}
\label{eq:QFIphi0}
    \mathcal{J}_\text{S}=M_1^2(1-2p^2)^2+4M_1p^2(1-p^2)\;,
\end{equation}
i.e.~the decoherence from this effect does not grow exponentially in the number of channel uses. An almost identical calculation follows for the GHZ state. As such, this effect will not change the value of $M_1$ that Heisenberg scaling can be achieved for. In general, however, this is a $\phi$-dependent effect. Using Refs.~\cite{dittmann1999explicit,liu2020quantum}, we can compute the QFI  analytically for any $M_1,p$, $\phi$, however the expression, while simple to compute, is too cumbersome to present here. In Fig.~\ref{fig:equivsyndrQFI}, we plot the QFI as a function of $\phi$ for fixed $M_1$ and $p$, demonstrating that in the regime of interest, Heisenberg scaling is still achieved in $M_1$.

A more convenient way to understand the effect of non-zero $\phi$ is as follows. We assume $-1/M_1\leq\phi\leq1/M_1$ (as is standard to avoid phase wrapping), and $M_1=O(1/\tilde{p}_{\text{L},N})=O(1/\phi)$. Then we can see that the off-diagonal terms in Eq.~\eqref{eq:apenB4rho} can be approximated as
\begin{equation}
\begin{split}
        &(\cos(\phi)+i(1-2p^2)\sin(\phi))^{c/\phi}\\
        \approx&\mathrm{e}^{ci(1-2p^2)}(1-2cp^2(1-p^2)\phi)\\
        =&\mathrm{e}^{ci(1-2p^2)}(1-2cp^2(1-p^2)O(\tilde{p}_\text{L}))\;.
\end{split}
\end{equation}
Hence, we can see that the total impact of this channel is to cause a $O(1)$ phase shift (simply the normal phase shift) and a decoherence that is not exponential in $M_1$. This effect therefore only changes the attainable precision by a constant factor. We can understand this effect as a dephasing error, with strength proportional to $O(p^2\phi)$. As this constant factor only enters at $O(p^2\tilde{p}_\text{L})$, we shall ignore this in our main results.

More generally, we have
\begin{equation}
\begin{split}
        &(\cos(\phi)+i(1-2p^2)\sin(\phi))^{M_1}\\
        \approx&\mathrm{e}^{M_1i(1-2p^2)\phi}(1-2M_1\phi^2p^2(1-p^2))\;.
\end{split}
\end{equation}
For any $M_1$ such that $|M_1\phi^2p^2|\ll1$, this dephasing error will be negligible. Equivalently, to not affect the main results we require $\phi^2p^2\ll O(\tilde{p}_\text{L})$.

\subsubsection{A discrete-time subtlety}
\label{apen:discretetime}
The above discussion highlights an important subtlety arising from the discrete-time model used---namely that above we modeled our channel as a bit-flip with probability $p$ followed by a rotation of one qubit. A more realistic channel should consider a total time $\delta t$ and the probability of a bit-flip should be equal across this whole time. In this setting, however, one can see that there is ambiguity in the true value of $\phi_{M_1}$, arising from the uncertainty as to when exactly in the window $\delta t$ did the bit-flip occur. This would create an error similar in magnitude to that above. Consider that the probability of a bit-flip occurring in time $\delta t$ is uniformly distributed across $\delta t$. For clarity we now distinguish between a bit-flip in the channel that occurs with probability $p_c$ and gate errors that occur with probability $p$. Ignoring gate errors for now, the phase acquired every round is with probability $1-p_c$ equal to $\phi$ and with probability $p_c$ drawn from a uniform distribution between $-\phi$ and $\phi$. Observe that 
\begin{equation}
    \frac{1}{2\phi}\int_{-\phi}^{\phi}\mathrm{e}^{-ix}dx=\frac{\sin(\phi)}{\phi}=\text{sinc}(\phi)\;.
\end{equation}
Then in one round (ignoring other errors for now), the GHZ coherences decay by a factor of $|1-p_c+p_c\mathrm{e}^{i\phi}\text{sinc}(\phi)|$. After $M_1$ rounds, the coherence will decay by a factor $|1-p_c+p_c\mathrm{e}^{i\phi}\text{sinc}(\phi)|^{M_1}$. For $\phi\ll1$ we have $1-p_c+p_c\mathrm{e}^{i\phi}\text{sinc}(\phi)\approx 1+ip_c\phi-2p_c\phi^2/3$. Therefore, $|1-p_c+p_c\mathrm{e}^{i\phi}\text{sinc}(\phi)|^2\approx 1-4p_c\phi^2/3$, and $|1-p_c+p_c\mathrm{e}^{i\phi}\text{sinc}(\phi)|\approx 1-2p_c\phi^2/3$ to leading order in $p_c$. This term would allow Heisenberg scaling for $M_1\approx1/(p_c\phi^2)$. Treating the gate errors separately, these errors impose the restriction $M_1\approx1/\tilde{c}_{N,t'}p^{t'}$. Hence, if $p_c\phi^2<\tilde{c}_{N,t'}p^{t'}$, the region over which we can achieve Heisenberg scaling will not be significantly reduced by this effect. Under the same assumption as above, $-1/M_1\leq\phi\leq1/M_1$, the results from the main text are unaffected.


If the experimenter can interact with the sensing channel for a time window $\delta t$, it is natural to assume $p_c\propto \delta t$. We will also have that $\phi\propto \delta t$. Hence $p_c\phi^2\propto \delta t^3$. By making $\delta t$ sufficiently small, we can always satisfy $p_c\phi^2<\tilde{c}_{N,t'}p^{t'}$, ensuring that channel errors are not dominant, and Heisenberg scaling is restored to arbitrary order. The total interaction time in this case is given by $T=\delta t M_1$, with $M_1$ restricted to be $M_1=O(1/\tilde{p}_{\text{L},N})$. This gives a total variance that scales as $1/T^2=O(\tilde{p}_{\text{L},N}^2/\delta t^2)$. To minimize the variance, we should therefore maximize $\delta t$ subject to the restriction $C (\delta t)^3<\tilde{p}_{\text{L},N}$, where $C$ is a constant. This implies we should just set $\delta t=O(\tilde{p}_{\text{L},N}^{1/3})$, giving a total variance of $O(\tilde{p}_{\text{L},N}^{4/3})$. Therefore, by increasing the code distance, the regime over which Heisenberg scaling is achieved can be increased with increasing code distance. We stress again that when $\phi$ is sufficiently small this effect is negligible and the results in the main text hold.



\subsection{Required measurement}
\label{apen:FTmeas}
 After the above steps, we have created the state given in Eq.~\eqref{eq:FT3qubitpremeas}. The final stage is to implement a measurement to extract the $\phi$ information while retaining Heisenberg scaling. In the ideal setting, the measurement to implement is discussed in Sec.~\ref{apen:problemreduction}. We now examine the effect of errors at every stage, focusing on the measurement that involves uncomputing the GHZ state and measuring only a single qubit. Assume we start from a perfect GHZ state for simplicity, $(\mathrm{e}^{-i\phi_{M_1}/2}\ket{0}^{\otimes 3}+\mathrm{e}^{i\phi_{M_1}/2}\ket{1}^{\otimes 3})/\sqrt{2}$. The first operation is a CNOT gate controlled on the second qubit targeting the third qubit, giving the state
 \begin{equation}
\frac{(\mathrm{e}^{-i\phi_{M_1}/2}\ket{0}^{\otimes 2}+\mathrm{e}^{i\phi_{M_1}/2}\ket{1}^{\otimes 2})\otimes \ket{0}}{\sqrt{2}}.
 \end{equation}
In our model, we then have $X$ errors everywhere giving the state (ignoring now the third qubit)
 \begin{widetext}
     \begin{equation}
     \begin{split}
\rho=&\frac{(1-p)^2}{2}(\mathrm{e}^{-i\phi_{M_1}/2}\ket{00}+\mathrm{e}^{i\phi_{M_1}/2}\ket{11})(\mathrm{e}^{i\phi_{M_1}/2}\bra{00}+\mathrm{e}^{-i\phi_{M_1}/2}\bra{11})\\
&+\frac{(1-p)p}{2}(\mathrm{e}^{-i\phi_{M_1}/2}\ket{10}+\mathrm{e}^{i\phi_{M_1}/2}\ket{01})(\mathrm{e}^{i\phi_{M_1}/2}\bra{10}+\mathrm{e}^{-i\phi_{M_1}/2}\bra{01})\\
&+\frac{(1-p)p}{2}(\mathrm{e}^{-i\phi_{M_1}/2}\ket{01}+\mathrm{e}^{i\phi_{M_1}/2}\ket{10})(\mathrm{e}^{i\phi_{M_1}/2}\bra{01}+\mathrm{e}^{-i\phi_{M_1}/2}\bra{10})\\
&+\frac{p^2}{2}(\mathrm{e}^{-i\phi_{M_1}/2}\ket{11}+\mathrm{e}^{i\phi_{M_1}/2}\ket{00})(\mathrm{e}^{i\phi_{M_1}/2}\bra{11}+\mathrm{e}^{-i\phi_{M_1}/2}\bra{00}).
     \end{split}
 \end{equation} 
 Applying the next CNOT operation, controlled on the first qubit, gives
     \begin{equation}
     \begin{split}
\rho=&\frac{(1-p)^2}{2}(\mathrm{e}^{-i\phi_{M_1}/2}\ket{00}+\mathrm{e}^{i\phi_{M_1}/2}\ket{10})(\mathrm{e}^{i\phi_{M_1}/2}\bra{00}+\mathrm{e}^{-i\phi_{M_1}/2}\bra{10})\\
&+\frac{(1-p)p}{2}(\mathrm{e}^{-i\phi_{M_1}/2}\ket{11}+\mathrm{e}^{i\phi_{M_1}/2}\ket{01})(\mathrm{e}^{i\phi_{M_1}/2}\bra{11}+\mathrm{e}^{-i\phi_{M_1}/2}\bra{01})\\
&+\frac{(1-p)p}{2}(\mathrm{e}^{-i\phi_{M_1}/2}\ket{01}+\mathrm{e}^{i\phi_{M_1}/2}\ket{11})(\mathrm{e}^{i\phi_{M_1}/2}\bra{01}+\mathrm{e}^{-i\phi_{M_1}/2}\bra{11})\\
&+\frac{p^2}{2}(\mathrm{e}^{-i\phi_{M_1}/2}\ket{10}+\mathrm{e}^{i\phi_{M_1}/2}\ket{00})(\mathrm{e}^{i\phi_{M_1}/2}\bra{10}+\mathrm{e}^{-i\phi_{M_1}/2}\bra{00})\;.
     \end{split}
 \end{equation} 
Discarding the second qubit gives
      \begin{equation}
     \begin{split}
\rho=&\frac{(1-p)^2+(1-p)p}{2}(\mathrm{e}^{-i\phi_{M_1}/2}\ket{0}+\mathrm{e}^{i\phi_{M_1}/2}\ket{1})(\mathrm{e}^{i\phi_{M_1}/2}\bra{0}+\mathrm{e}^{-i\phi_{M_1}/2}\bra{1})\\
&+\frac{p^2+(1-p)p}{2}(\mathrm{e}^{-i\phi_{M_1}/2}\ket{1}+\mathrm{e}^{i\phi_{M_1}/2}\ket{0})(\mathrm{e}^{i\phi_{M_1}/2}\bra{1}+\mathrm{e}^{-i\phi_{M_1}/2}\bra{0})\;.
     \end{split}
 \end{equation} 
  The next round of bit-flip errors then gives
        \begin{equation}
     \begin{split}
\rho=&\frac{(1-p)((1-p)^2+(1-p)p)+p(p^2+(1-p)p)}{2}(\mathrm{e}^{-i\phi_{M_1}/2}\ket{0}+\mathrm{e}^{i\phi_{M_1}/2}\ket{1})(\mathrm{e}^{i\phi_{M_1}/2}\bra{0}+\mathrm{e}^{-i\phi_{M_1}/2}\bra{1})\\
&+\frac{(1-p)(p^2+(1-p)p)+p((1-p)^2+(1-p)p)}{2}(\mathrm{e}^{-i\phi_{M_1}/2}\ket{1}+\mathrm{e}^{i\phi_{M_1}/2}\ket{0})(\mathrm{e}^{i\phi_{M_1}/2}\bra{1}+\mathrm{e}^{-i\phi_{M_1}/2}\bra{0})\\
=&\frac{1-2p(1-p)}{2}(\mathrm{e}^{-i\phi_{M_1}/2}\ket{0}+\mathrm{e}^{i\phi_{M_1}/2}\ket{1})(\mathrm{e}^{i\phi_{M_1}/2}\bra{0}+\mathrm{e}^{-i\phi_{M_1}/2}\bra{1})\\
&+\frac{2(1-p)p}{2}(\mathrm{e}^{-i\phi_{M_1}/2}\ket{1}+\mathrm{e}^{i\phi_{M_1}/2}\ket{0})(\mathrm{e}^{i\phi_{M_1}/2}\bra{1}+\mathrm{e}^{-i\phi_{M_1}/2}\bra{0})\;.
     \end{split}
 \end{equation} 
All of this is to say that, before the final Hadamard in Fig.~\ref{fig:FTmeas}, the single qubit of interest experiences an even number of errors with probability $(1-p)^2 +p^2$ (a factor $1-p$ each from idling, and the CNOT). The probability of an odd number of errors is therefore $2p(1-p)$, giving the state
 \begin{equation}
    ( 1-2p(1-p))\ket{\psi(\phi_{M_1})}\bra{\psi(\phi_{M_1})}+(2p(1-p))\ket{\psi(-\phi_{M_1})}\bra{\psi(-\phi_{M_1})}\;.
 \end{equation}
  \end{widetext}
Next we act with a Hadamard gate, giving the state
\begin{equation}
\rho=\begin{pmatrix}
     \cos ^2\left(\frac{\phi_{M_1} }{2}\right) & \frac{i}{2}  (1-2 p)^2 \sin (\phi_{M_1} ) \\
 -\frac{i}{2} (1-2 p)^2 \sin (\phi_{M_1} ) & \sin ^2\left(\frac{\phi_{M_1} }{2}\right) \\
\end{pmatrix}\;.
\end{equation}
This demonstrates a slightly surprising effect---if we were to introduce no further errors (the bit-flip following the Hadamard gate still needs to be accounted for), then the previous errors acquired during the uncomputation of the GHZ state would not degrade the precision, i.e. bit-flip errors after phase accumulation do not affect precision, contrary to those during phase accumulation (Sec.~\ref{subsec:signoisediff}). After the Hadamard, because of the basis change, bit-flip errors act like $Z$ errors and are now detrimental to precision. The final state before measuring (after the final round of bit-flip errors) is
\begin{equation}
    \rho=\begin{pmatrix}
     \frac{1}{2}+\frac{1}{2}\cos \left(\phi_{M_1}\right)(1-2p) & \frac{i}{2}  (1-2 p)^3 \sin (\phi_{M_1} ) \\
 -\frac{i}{2} (1-2 p)^3 \sin (\phi_{M_1} ) & \frac{1}{2}-\frac{1}{2}\cos \left(\phi_{M_1}\right)(1-2p) \\
\end{pmatrix}\;.
\end{equation}
Measuring in the computational basis, we find the probability of each outcome is given by
 \begin{equation}
     p_\pm=\frac{1}{2}(1\pm(1-2p)\cos(\phi_{M_1}))\;.
 \end{equation}
 Accounting for a measurement bit-flip with probability $p_m$, we arrive at the following probabilities:
 \begin{equation}
        p_\pm=\frac{1}{2}(1\pm(1-2p_m)(1-2p)\cos(\phi_{M_1}))\;. 
 \end{equation}
 The CFI can then be computed as (using $\langle\phi_{M_1}\rangle=M_1\phi(1-2p)$)
\begin{equation}
     \mathcal{J}=-\frac{M_1^2 (1-2 p)^4 (1-2 p_m)^2 \sin ^2(\phi_{M_1} )}{(1-2 p)^2 (1-2 p_m)^2 \cos ^2(\phi_{M_1} )-1}\;.
\end{equation}
As above, by choosing the operating point appropriately (e.g.~via an adaptive protocol, or via a single $R_z$ rotation applied at the end of the protocol), this achieves Heisenberg scaling. Operating at $\phi=\pi/(2M_1(1-2p))$, shows that compared to the ideal case, imperfections in the measurement reduce the CFI by a factor $(1-2 p)^2(1-2 p_m)^2$.

\subsection{$N$-qubit repetition code}
\label{apen:nqubitrepetition}
To analyze the $N$-qubit repetition code in more detail, we use $p$, $p_g$, and $p_m$ to denote the probability of a physical error occurring during the sensing channel, after every gate, idling or state preparation operation, and after a measurement,  respectively. For an $N$-qubit repetition code, after state preparation, we have the state
\begin{equation}
    \rho_{\text{GHZ},N}(p_g)=(1-\tilde{p}_{\text{L,p},N})\ket{\psi_\text{GHZ}}\bra{\psi_\text{GHZ}}+O(\tilde{p}_{\text{L,p},N})\;,
\end{equation}
where $\tilde{p}_{\text{L,p},N}$ is the probability of a logical error during state preparation, and is a function of both $p_g$ and $p_m$. The $N$-qubit repetition code can correct $t=\lfloor (N-1)/2\rfloor$ errors. As is standard, if $p_m\approx p_g$, then $\tilde{p}_{\text{L,p},N}=O(p_g^{t'})$, where $t'=t+1$. We now prove this for the $N$-qubit repetition code. According to our error model, when a qubit is initialized in the $\ket{0}$ state, the state $\ket{1}$ is prepared with probability $p_g$. After a Hadamard gate acts on this state, the states $\ket{+}$ and $\ket{-}$ are prepared with probability $1-p_g$ and $p_g$ respectively. An ancilla state is then used to verify the correct state was prepared as shown in Fig.~\ref{fig:FTstateprep}. The CNOT gate between the $\ket{\pm}$ state and the ancilla introduces only $X$ errors which do not affect the state $\ket{\pm}$ (up to an irrelevant phase). Other errors on the ancilla qubit affect the measurement result but do not affect the data qubit. By repeating the process of measuring $\langle X \rangle$ sufficiently many times (exactly how many times are required is discussed in Sec.~\ref{apensyndromerepetitions}), the probability of such measurement errors can be arbitrarily reduced. This enables the $\ket{+}$ state to be prepared with probability $1-O(\tilde{p}_{\text{L,p},N})$. The second stage in preparing a GHZ state fault-tolerantly is a fan-out CNOT circuit as shown in Fig.~\ref{fig:FTstateprep}. An $X$ error during this step can propagate through the chain of CNOTs, however this does not cause an uncorrectable logical error. This is because, for odd $N=2t+1$,
\begin{equation}
    X^{\otimes N}\ket{\psi_{\text{GHZ}}}
=\ket{\psi_{\text{GHZ}}}\;,
\end{equation}
so every error $X^S$ is equivalent on the target state to a different error
$X^{S^c}=X^S X^{\otimes N}$. Since
$\min\{|S|,N-|S|\}\le t$, every resulting $X$-error pattern is
equivalent to a correctable error. Thus, the fan-out circuit cannot
produce a logical state-preparation error under the assumed $X$-only
noise model. All that remains is to show that these errors can be detected and corrected in a fault-tolerant manner. Errors are detected by measuring $Z_iZ_{i+1}$, as shown in Fig.~\ref{fig:FTstateprep}. For the CNOT operations involved in this measurement, the data qubits are the control and the ancilla is the target. As above, under our noise model, this error detection procedure cannot cause hook errors: a single fault in a stabilizer-measurement circuit produces at most one data-qubit error together with a local measurement error, and cannot generate a correlated multi-data-qubit hook error. Therefore, by repeating the syndrome measurement at least $2t+1$ times (see Sec.~\ref{apensyndromerepetitions}) we obtain a space-time detector graph of distance at least $2t+1$. Since each individual preparation, gate, or measurement error produces at most one local edge in this graph, every pattern of at most $t$ faults is correctly decoded giving rise to the claimed scaling. The corresponding correction operation can be implemented using only single qubit $X$ operations and so does not affect this claim.

This state then interacts with the signal and noise channel, and we attempt to extract information about errors in the channel using a noisy syndrome extraction procedure, that involves $r$ repeated measurements of each syndrome. Correcting $t$ physical errors, logical errors can occur in syndrome extraction with probability
\begin{equation}
\tilde{p}_{\text{L},N}=  \sum_{i,j,k\geq0, i+j+k=t'}c_{i,j,k}(r)p_g^ip^jp_m^k+O(\max{(p_g,p,p_m)}^{t'+1})\;, 
\end{equation}
where $c_{i,j,k}(r)$ is a constant that counts the number of ways in which $i$ gate errors, $j$ channel errors, and $k$ measurement errors can cause a logical error. The syndrome extraction procedure above succeeds with probability $1-\tilde{p}_{\text{L},N}$. Hence, with probability $1-\tilde{p}_{\text{L},N}$, errors are suppressed to $p^{t'}$. Equivalently, with probability $\tilde{p}_{\text{L},N}$, logical errors occur. After $M_1$ rounds, the probability of no logical errors is $(1-\tilde{p}_{\text{L},N})^{M_1}$. Proving this scaling proceeds in a similar manner to above. Errors are detected by measuring $Z_iZ_{i+1}$ repeatedly and the complete syndrome history
is decoded using minimum-weight perfect matching. As above, the syndromes $Z_iZ_{i+1}$ are measured using CNOT gates oriented such that the data qubits are the control and the ancilla qubit is the target. This causes only local $X$ errors on data qubits $i$ and $i+1$, and possible errors on the ancilla qubit used to measure the syndrome. Thus, a single channel, gate, or measurement error produces at most one local edge in the spacetime
decoding graph and cannot generate a correlated data-qubit hook error. Again, repeating the syndrome measurements at least $2t+1$ times gives a space-time decoding distance of $2t+1$, and so $t+1$ errors are required to cause a logical error.

It only remains to examine the measurement required to achieve Heisenberg scaling. We again consider the measurement whereby the GHZ state is uncomputed and all of the measurement information is extracted via a single-qubit measurement. An $N$-qubit GHZ state can be uncomputed with $r_N=\lceil N/2 \rceil$ rounds of CNOT gates (assuming a linear-connectivity circuit). In these rounds, the probability of an odd number of bit-flip errors happening on the single qubit of interest is 
\begin{equation}
    p_x=\frac{1-(1-2p_g)^{r_N}}{2}\;.
\end{equation}
The single-qubit state containing information about $\phi$ is then
\begin{widetext}
    \begin{equation}
    ( 1-p_x)\ket{\psi(\phi_{M_1})}\bra{\psi(\phi_{M_1})}+p_x\ket{\psi(-\phi_{M_1})}\bra{\psi(-\phi_{M_1})}\;.
 \end{equation}
Next we act with a Hadamard gate, giving the state
\begin{equation}
\rho=\begin{pmatrix}
     \cos ^2\left(\frac{\phi_{M_1} }{2}\right) & \frac{i}{2}  (1-2 p_x) \sin (\phi_{M_1} ) \\
 -\frac{i}{2} (1-2 p_x) \sin (\phi_{M_1} ) & \sin ^2\left(\frac{\phi_{M_1} }{2}\right) \\
\end{pmatrix}\;.
\end{equation}
Accounting for bit-flips after the Hadamard, we arrive at
\begin{equation}
    \rho=\begin{pmatrix}
     \frac{1}{2}+\frac{1}{2}\cos \left(\phi_{M_1}\right)(1-2p_g) & \frac{i}{2}  (1-2 p_g)(1-2p_x) \sin (\phi_{M_1} ) \\
 -\frac{i}{2} (1-2 p_g)(1-2p_x) \sin (\phi_{M_1} ) & \frac{1}{2}-\frac{1}{2}\cos \left(\phi_{M_1}\right)(1-2p_g) \\
\end{pmatrix}\;.
\end{equation}
Remarkably $p_x$ does not enter into the CFI. Hence, even in the $N$ qubit setting, the effect of a non-fault-tolerant measurement is simply to reduce the CFI by a factor $(1-2p_g)^2(1-2p_m)^2$.\footnote{ We note that measuring the logical $X$ operator can be done in a fault-tolerant manner potentially reducing these factors. This can be done in a standard way~\cite{tomita2013comparison}, with a variant of this adapted in Ref.~\cite{sahu2026achieving}.} Hence, the CFI is lower bounded by
\begin{equation}
    \mathcal{J}\geq(2(1-\tilde{p}_{\text{L},\text{p},N})(1-\tilde{p}_{\text{L},N})^{M_1}-1)^2
    (1-2p_m)^2(1-2p_g)^2(1-2p)^2M_1^2\;.
\end{equation}
This is maximized when $M_1\approx-0.315/\log(1-\tilde{p}_{\text{L},N})\approx 0.315/\tilde{p}_{\text{L},N}=O(1/p^{t'})$. However, the overall variance given $M=M_1\nu$ total channel uses is minimized when $M_1\approx-0.2/\log(1-\tilde{p}_{\text{L},N})\approx a/\tilde{p}_{\text{L},N}=O(1/p^{t'})$ for $a=0.2$. For simplicity, assume $p_g=p_m=p$, then
    \begin{equation}
\begin{split}
      \mathcal{J}&\geq(2(1-\tilde{p}_{\text{L},\text{p},N})(1-\tilde{p}_{\text{L},N})^{M_1}-1)^2
    (1-2p)^6M_1^2\\
      &\approx\frac{(2(1-\tilde{p}_{\text{L},\text{p},N})(1-\tilde{p}_{\text{L},N})^{a/\tilde{p}_{\text{L},N}}-1)^2
    (1-2p)^6a^2}{\tilde{p}_{\text{L},N}^2}\\
     &\approx\frac{(2(1-\tilde{p}_{\text{L},\text{p},N})\mathrm{e}^{-a}-1)^2
    (1-2p)^6a^2}{\tilde{p}_{\text{L},N}^2}\\
    &\approx\frac{
    (1-2p)^6}{60\tilde{p}_{\text{L},N}^2}\\
\end{split}
\end{equation}
\end{widetext}
using $M_1=a/\tilde{p}_{\text{L},N}$ with $a=0.2$.
This demonstrates that the range over which Heisenberg scaling can be achieved can be arbitrarily improved with larger codes. The corresponding variance bound is
\begin{equation}
    \sigma_{N,\phi}^2\leq\frac{1}{(2(1-\tilde{p}_{\text{L},\text{p},N})(1-\tilde{p}_{\text{L},N})^{M_1}-1)^2
    (1-2p)^6M_1^2\nu}\;,
\end{equation}
valid when $(1-\tilde{p}_{\text{L},\text{p},N})(1-\tilde{p}_{\text{L},N})^{M_1}>1/2$. We emphasize that the region in which this equation is valid is precisely the region of interest, as the improvement in performance from fault-tolerant quantum sensing is greatest when \mbox{$p\ll1$}, and therefore \mbox{$\tilde{p}_{\text{L},N}\ll1$}.

\subsubsection{How many syndrome repetitions are needed?}
\label{apensyndromerepetitions}
Finally, it is worth discussing how many times the syndrome measurements need to be repeated to achieve the desired error in the measured syndrome value. Note that here we analyse only the syndrome-\emph{decision} error, while data and gate faults are handled by the spacetime matching argument in appendix~\ref{apen:nqubitrepetition}. In each round, we measure each stabilizer $r$ times using an identical noisy measurement circuit, and then decode the check value from the $r$ outcomes.
If a single measurement is wrong with probability $\eta$ (assuming $\eta=O(p_g)+O(p_m)$, absorbing the number of fault locations in that check circuit into the constant), then under a simple majority-vote decoder (with $r$ odd) the probability of decoding the syndrome incorrectly is
\begin{equation}
\begin{split}
    P_{\mathrm{mv}}(r)
&=\sum_{m=(r+1)/2}^{r}\binom{r}{m}\eta^{m}(1-\eta)^{r-m}
\\
&\le\;\binom{r}{(r+1)/2}\eta^{(r+1)/2},
\end{split}
\label{eq:majority_vote_bound}
\end{equation}
where the last inequality holds for $\eta\le 1/2$.
Thus majority voting boosts the effective reliability of the syndrome measurement result from $O(\eta)$ to $O(\eta^{(r+1)/2})$.

For a $N$-qubit repetition code correcting $t=\lfloor(N-1)/2\rfloor$ faults per round, it is sufficient to choose $r$ so that the probability of an incorrectly decoded syndrome is at most the target logical order,
\begin{equation}
P_{\mathrm{mv}}(r)=O(\tilde{p}_{\text{L},N}),
\label{eq:r_target_order}
\end{equation}
so that syndrome-decision failures do not dominate the leading logical error scaling.
Since $\eta=O(p_g+p_m)$, Eq.~\eqref{eq:majority_vote_bound} implies that it suffices to take
\begin{equation}
\frac{r+1}{2}\;\ge\;t+1
\qquad\Longleftrightarrow\qquad
r\;\ge\;2t+1,
\label{eq:r_scaling}
\end{equation}
up to constant-factor improvements that depend on the check circuit and decoder.

It is worth noting that above we have modelled the $r$ outcomes as independent and identically distributed Bernoulli random variables, each equal to the incorrect check value with probability $\eta$, conditioned on a fixed underlying true check value. While this is correct for readout and ancilla faults, it is not the correct model for a data fault occurring midway through the $r$ repetitions, since that changes the true value being voted on. However, this case is precisely what the spacetime decoder in appendix~\ref{apen:nqubitrepetition} handles.

\subsection{Effect of phase-flip errors}
\label{apen:zerrors}
Because the repetition code corrects $X$-type faults but does not detect $Z$ faults, adding dephasing (random $Z$ operators) with probability $p_z$ per round introduces an additional logical phase-flip channel that suppresses the GHZ coherence by a factor $(1-2p_z)^{M_1}$. (We stress that $p_z$ is the probability of a $Z$-type error anywhere in the circuit causing a logical phase-flip, not the probability of an error on a single qubit.) The $Z$ errors would cause additional decoherence, for example in the state preparation and measurement, but to leading order the CFI will be reduced by $(1-2p_z)^{2M_1}$. If $p_z=O(\tilde{p}_{\text{L},N})$, this does not affect the Heisenberg scaling. Note however, that the required bias between phase-flip and bit-flip rates must itself scale with the code distance, and a fixed experimental bias will eventually be insufficient at large $N$.

\section{Heisenberg scaling for qudit systems}
\label{apen:qudit}
In the main text, we used a standard error correcting code (the repetition code) to achieve Heisenberg scaling for qubit systems under a restricted noise model. In this appendix, we study a qudit version of this.

\subsection{Qutrit example}
We first study a qutrit example, starting with the generalized operators $Z\ket{k}=\omega^k\ket{k}$ ($\omega=\mathrm{e}^{i2\pi/3}$) and $X\ket{k}=\ket{k+1 \mod(3)}$:
\begin{equation}
    \begin{split}
        Z=&\begin{pmatrix}
            1&0&0\\
            0&\omega&0\\
            0&0&\omega^2
        \end{pmatrix},
        \\
    X=&\begin{pmatrix}
            0&0&1\\
            1&0&0\\
            0&1&0        \end{pmatrix}.
    \end{split}
\end{equation}
Additionally we will need the following controlled operations $C_{3,+}\ket{x,y}\to\ket{x,y+x \mod (3)}$ and $C_{3,-}\ket{x,y}\to\ket{x,y-x \mod (3)}$. The Hamiltonian we study acts only on the first qudit, and is given by
\begin{equation}
    H_1=\phi(Z_1+Z_1^\dagger)/2\;,
\end{equation}
and the allowed channel errors are $X$ and $X^2$. One can verify that the HNLS condition is satisfied for this problem: the span of the jump operators is $\mathrm{span}\{I,X,X^2\}$. Every operator in this span is circulant, whereas the signal generator
$h_0=\operatorname{diag}(1,-1/2,-1/2)$ is a non-scalar diagonal matrix. Hence, $h_0\notin\mathcal S$, and the
HNLS condition is satisfied.
We first study a simple error correction procedure with errors only during the evolution of the probe state before extending this to a fully fault-tolerant protocol, capable of handling $X$ and $X^2$ errors everywhere.

\subsubsection{Qutrit error-corrected sensing}
The optimal sensing probe state without any noise is
\begin{equation}
    \ket{\psi}=\frac{1}{\sqrt{2}}(\ket{0}+\ket{2})\;.
\end{equation}
However, to correct errors in the channel, we introduce a state entangled across ancilla qubits
\begin{equation}
    \ket{\psi}=\frac{1}{\sqrt{2}}(\ket{0}^{\otimes N}+\ket{2}^{\otimes N})\;.
\end{equation}
Under ideal evolution, this would become
\begin{equation}
        \ket{\psi}=\frac{1}{\sqrt{2}}(\mathrm{e}^{-i\phi}\ket{0}^{\otimes N}+\mathrm{e}^{i\phi/2}\ket{2}^{\otimes N})\;.
\end{equation}
$X$ or $X^2$ errors can move the state out of this space, and they do so in fundamentally different ways---allowing them to be distinguished. For example an $X$ error changes $\ket{0}+\ket{2}$ to $\ket{1}+\ket{0}$, whereas an $X^2$ error changes $\ket{0}+\ket{2}$ to $\ket{2}+\ket{1}$. After each round of measurement, up to $t=\lfloor(N-1)/2\rfloor$ errors can be corrected, which gives Heisenberg scaling for a number of channel repetitions that scales as $1/p^{t'}$. This is a standard procedure following directly from the fact that the HNLS condition from Ref.~\cite{zhou2018achieving} is satisfied.

One way of carrying out the syndrome extraction is by measuring the stabilizer $Z_iZ_{i+1}^\dagger$. This is done by applying $C_{3,+}$ controlled on qubit $i$ to an ancilla target, followed by $C_{3,-}$ from qubit $i+1$ to the ancilla. After this operation, measuring the ancilla gives $Z_iZ_{i+1}^\dagger$. As an example consider the code $(\ket{0}^{\otimes 3}+\ket{2}^{\otimes 3})/\sqrt{2}$. An $X$ error on the first qubit transforms this to $(\ket{100}+\ket{022})/\sqrt{2}$. Hence after acting $C_{3,+}$ and $C_{3,-}$, the joint state with ancilla is $\ket{1001}+\ket{0221}$. Similarly, after an $X^2$ error on the first qubit the system is in the state $(\ket{200}+\ket{122})/\sqrt{2}$. After the control operations the joint system is $(\ket{2002}+\ket{1222})/\sqrt{2}$. And so the specific error that occurred can be determined. More generally, by measuring all $Z_iZ_{i+1}^\dagger$, the location of the error can be determined and corrected.

\subsubsection{Qutrit fault-tolerant sensing}
We now demonstrate how the repetition code from above with repeated syndrome extractions can be used for fault-tolerant sensing. We consider the restricted noise model such that, after every gate, state preparation, or idling sequence, the operators $X$ and $X^2$ act on each qutrit with probability $p$. For state preparation, we use an approach similar to that in Fig.~\ref{fig:FTstateprep}. We use the same control operations as above, i.e.~measure the stabilizer $Z_iZ_{i+1}^\dagger$, by applying $C_{3,+}$ controlled on qubit $i$ to an ancilla target, followed by $C_{3,-}$ from qubit $i+1$ to the ancilla. To make the syndrome information reliable under circuit-level faults, we repeat the stabilizer-measurement round $r=\Theta(t)$ times (e.g.\ $r=2t+1$) and decode the resulting space--time syndrome, so that the probability of a wrong decoded syndrome is suppressed to $O(p^{t+1})$ for a distance-$(2t+1)$ repetition code. Fault-tolerant syndrome extraction and the final measurement proceed similarly, giving the same result as before. The major additional assumption is that the $C_{3,\pm}$ operator can be implemented while only introducing $X$ or $X^2$ errors. This is a natural analogue of the qubit assumptions made in the main text.

\subsection{Generalized qudit example}
The above example can be generalized to the $D$-dimensional case with the operators $Z\ket{k}=\omega^k\ket{k}$ ($\omega=\mathrm{e}^{i2\pi/D}$), $X\ket{k}=\ket{k+1 \mod(D)}$, $C_{D,+}\ket{x,y}\to\ket{x,y+x\mod (D)}$, and $C_{D,-}\ket{x,y}\to\ket{x,y-x\mod (D)}$. The Hamiltonian again acts only on the first qudit and is \mbox{$H_1=\phi h_0=\phi(Z_1+Z_1^\dagger)/2$}. The eigenvalues of $h_0$ are $\lambda_k=\cos(2\pi k/D)$, and so the probe state used is $
\ket{\psi_L}
=
\frac{\ket{0}^{\otimes N}+\ket{m}^{\otimes N}}{\sqrt2}
$, where $m=\lfloor D/2\rfloor$. After evolution for time $t$, this state picks up a relative phase of $(\lambda_0-\lambda_m)\phi t$.

 For the qudit problem, the errors allowed during any operation are $X,X^2,...X^{D-1}$. Similar to above, the HNLS condition is satisfied under this noise model. Stabilizers are then measured in a similar way allowing the above class of errors to be detected.

\subsection{General Case}
The HNLS condition characterizes when Heisenberg scaling is achievable under an ideal recovery channel for a fixed sensing Lindbladian.
To extend this to circuit-level fault tolerance, one must additionally specify a fault model for state preparation, syndrome extraction, and measurement.
A natural sufficient framework is the existence of (i) a family of codes $\{\Pi\}$ correcting up to $t$ errors from a restricted, signal-avoiding error set, (ii) a signal that acts non-trivially on these code words, and (iii) syndrome-extraction and recovery gadgets that allow the construction of a space-time detector graph of distance at least $2t+1$.

Under these assumptions, the probability of an uncorrectable event per correction cycle can be suppressed from $O(p)$ to $O(p^{t+1})$, extending the Heisenberg-limited interrogation window accordingly.
Our qubit and qudit repetition-code constructions realize this program for restricted error models, while extending the construction to general HNLS codes remains an open problem.

While HNLS guarantees the existence of a code and recovery achieving Heisenberg scaling under idealized (noiseless, fast) control, it does not imply the existence of a fault-tolerant circuit implementing that recovery: circuit-level faults can enlarge the effective error set through propagation or time-dependent backaction, and implementing the required logical signal evolution may demand non-transversal continuous rotations with prohibitive overhead. Even in the simplest qubit example above, we require the restriction that all gates introduce only $X$ errors. This by itself is quite artificial. 

\section{Worst-case bound over logical phase histories}
\label{apenworstcase}

Let
\begin{equation}
q=(1-\widetilde p_{\mathrm{L,p},N})
  (1-\widetilde p_{\mathrm L,N})^{M_1}    
\end{equation}
denote the probability of the successfully corrected branch. Under the
$X$-only noise model, each fault history leaves the encoded state,
up to a correctable error, in the logical GHZ subspace with an
accumulated phase $k\phi$, where $k\in[-M_1,M_1]$. We therefore write the density matrix of our final state as
\begin{equation}
\begin{split}
      \rho(\phi)
=&
q\ket{\psi_\text{GHZ}(M_1\phi)}\bra{\psi_\text{GHZ}(M_1\phi)}\\&
+
(1-q)\sum_h w_h
\ket{\psi_\text{GHZ}(k_h\phi)}\bra{\psi_\text{GHZ}(k_h\phi)},
\end{split}
\end{equation}
where $w_h\geq0$, $\sum_h w_h=1$, $k_h$ is an integer of the same parity as $M_1$, $|k_h|\leq M_1$, and the fault-history
probabilities are independent of $\phi$.

We consider a logical-$Y$ measurement with outcome $x=\pm1$. The mean of such a measurement is
\begin{equation}
    m(\phi)
=
q\sin(M_1\phi)
+
(1-q)\sum_h w_h\sin(k_h\phi).
\end{equation}
Near $\phi=0$, $m(0)=0$ and
\begin{equation}
\begin{split}
    m'(0)&
=
qM_1+(1-q)\sum_h w_h k_h
\\&\geq
qM_1-(1-q)M_1
=
(2q-1)M_1.
\end{split}
\end{equation}
As the measurement in the $Y$ basis gives a binomial distribution with probabilities
\begin{equation}
P(x|\phi)=\frac12[1+xm(\phi)]\;,
\end{equation}
the corresponding CFI at $\phi=0$ is
\begin{equation}
    \mathcal J
=\frac{[m'(0)]^2}{1-m(0)^2}=
[m'(0)]^2
\geq
(2q-1)^2M_1^2\;.
\end{equation}
Consequently, for $q\geq1/2$, the $Y$-basis
measurement retains a Fisher information proportional to $M_1^2$ even
for the worst allowed distribution of logical phase histories.

\bibliography{ECbib}
\bibliographystyle{naturemag}

\end{document}